# Inverse design of Non-parameterized Ventilated Acoustic Resonator via Variational Autoencoder with Acoustic Response-encoded Latent Space

Min Woo Cho[1][*], Seok Hyeon Hwang[1][*], Jun Young Jang[1][*], Jin Yeong Song[1], Sunkwang Hwang[2], Kyoung Je Cha[2], Dong Yong Park[2][**], Kyungjun Song[1][**], Sang Min Park[1][**]

[1]*School of Mechanical Engineering, Pusan National University, 63-2 Busan University-ro, Geumjeong-gu, Busan, 46241, South Korea*

[2]*Smart Manufacturing Technology R&D Group, Korea Institute of Industrial Technology, 320 Techno sunhwan-ro, Yuga-eup, Dalseong-gun, Daegu, 42994, South Korea*

[*]*Equally contributed*

**Corresponding author:

sangmin.park@pusan.ac.kr (Sang Min Park)

dypark9606@kitech.re.kr (Dong Yong Park)

song3396@pusan.ac.kr (Kyungjun Song)

# Abstract

Ventilated acoustic resonator(VAR), a type of acoustic metamaterial, emerge as an alternative for sound attenuation in environments that require ventilation, owing to its excellent low-frequency attenuation performance and flexible shape adaptability. However, due to the non-linear acoustic responses of VARs, the VAR designs are generally obtained within a limited parametrized design space, and the design relies on the iteration of the numerical simulation which consumes a considerable amount of computational time and resources. This paper proposes an acoustic response-encoded variational autoencoder (AR-VAE), a novel variational autoencoder-based generative design model for the efficient and accurate inverse design of VAR even with non-parametrized designs. The AR-VAE matches the high-dimensional acoustic response with the VAR cross-section image in the dimension-reduced latent space, which enables the AR-VAE to generate various non-parametrized VAR cross-section images with the target acoustic response. AR-VAE generates non-parameterized VARs from target acoustic responses, which show a 25-fold reduction in mean squared error compared to conventional deep learning-based parameter searching methods while exhibiting lower average mean squared error and peak frequency variance. By combining the inverse-designed VARs by AR-VAE, multi-cavity VAR was devised for broadband and multitarget peak frequency attenuation. The proposed design method presents a new approach for structural inverse-design with a high-dimensional non-linear physical response.

# 1. Introduction

Achieving effective sound insulation in environments that require ventilation has been a longstanding challenge in the field of acoustics. To attain a high level of soundproofing, it is generally essential to physically block any gaps and openings that could allow sound waves to pass through and transmit acoustic energy. However, when ventilation is required, traditional soundproofing methods can lead to the deterioration of the ventilation system. This trade-off between sound insulation and ventilation has been observed in various sound-blocking apparatuses, such as vehicle mufflers [1-3], acoustic vents [4-6], and acoustic windows [7-10].

To address this challenge, acoustic metamaterials (AMs) have been proposed as a solution. AMs consist of sub-wavelength elements arranged in a repeating pattern. Such structures are designed to have extraordinary acoustic properties, such as zero/extreme/negative density [11-13] or negative bulk modulus [14-16], which enable them to control sound waves in ways that were previously considered impossible. Moreover, by utilizing these properties, they can be employed in open spaces to facilitate airflow, allowing the simultaneous control of mass transportation and acoustic energy. Among these designs, the ventilated acoustic resonator (VAR) has gained considerable attention owing to its simple design and adaptability to various shapes [17-22]. A VAR consists of two parts: a waveguide and a Helmholtz resonator. Because the waveguide allows fluid to flow and the Helmholtz resonator blocks noise, both mass transportation and acoustic energy can be controlled. In addition, the geometric parameters of each part can be easily adjusted for ventilation and soundproofing performance, offering a high degree of flexibility in configuring the system to meet specific engineering needs.

Considering the wide applications of VAR for sound attenuation, it is necessary to develop an inverse-design technology for VAR structures that reflect the acoustic characteristics of the target application. Inverse design, which searches for the optimal design with characteristics that meet the target performance, has been widely studied in the fields of photonics [23], aerodynamics [24], and acoustics [25], in which the relationships between the structure and dynamics are complex. In general, a VAR with a simple geometric shape that could be grasped by human intuition has been inverse-designed by analytical methods such as the transfer matrix method [26, 27]. However, as the geometrical complexity increases to achieve superior acoustic performance, the inverse design of VAR structures with the target acoustic response by human intuition or analytical methods would become increasingly challenging. Thus, inverse-design methods based on the iteration of numerical simulations such as genetic algorithms [28, 29], space-filling [30], direct binary search [31], and topology optimization [32] are widely used. However, these methods consume a significant amount of computational resources and time. Moreover, the optimal VAR structure that matches the target acoustic response could be omitted when the degree of freedom in the design space is high.

Deep learning, a field of machine learning that utilizes an artificial neural network that mimics the human nervous system, has garnered growing interest as a means of addressing intricate problems that are challenging to resolve through conventional methods, such as computer vision [33, 34], natural language processing [35], active control [36-38], and speech recognition [39]. In the field of structural design such as electromagnetic metamaterials [40-42], mechanical metamaterials [43, 44], composites [45], and microwave absorbers [46], numerous studies have been conducted on the inverse design of non-intuitive structures utilizing the characteristics of deep learning models. Furthermore, the design of acoustic structures, such as phononic crystals [47], low-frequency acoustic absorbers [48], acoustic sinks [25], acoustic scatterers [49], and acoustic metasurfaces [50-52] has also been subject to the deep learning model. However, previous studies majorly relied on deep learning-based parameter searching methods, which limited the degree of freedom of design space and corresponding acoustic response. Thus, it is required to develop a deep learning model for the inverse design of VAR with a higher degree of freedom in the design space while reflecting the target acoustic response.

In this study, we propose an acoustic response-encoded variational autoencoder (AR-VAE) for the accurate and inverse design of non-parameterized VAR, which is based on a deep generative model. A deep generative model, which belongs to the category of deep learning models, is designed to generate data with a similar distribution to the training dataset. Representative deep generative models include a generative adversarial network (GAN) [53] which utilizes the competing relationship between a discriminator network and generator network, and a variational autoencoder (VAE) [54] which uses a probabilistic approach. Unlike GAN, which more focuses on generating similar images to the training dataset, VAE learns the distribution of the training dataset explicitly. Furthermore, the dimension-reduced latent space of VAE could encode the geometrical features and acoustic response of the VAR, which is required for the accurate and precise inverse design. Therefore, in this study, we adopted VAE as a base model which is more suitable for understanding the distribution and features of VAR's geometric shape. We modified the VAE into the AR-VAE to encode a high-dimensional target acoustic response and generate non-parametrized VAR cross-section images with an acoustic response similar to the target acoustic response with high accuracy. To compare the inverse-design performance of AR-VAE with the inverse-design performance of the deep learning-based parameter searching method[25, 47, 48, 55-57] we trained an acoustic response-parameter matching neural network (APNN), a deep learning model that can search the geometric parameters of the VAR corresponds to the input acoustic response as a control group. We could confirm that the inverse-designed VAR by AR-VAE exhibits a 25-fold reduction in mean squared error with a target acoustic response compared to the inverse-designed VAR by APNN. As another validation, we parametrized the non-parametrized inverse-designed VAR and found that the non-parametrized inverse-designed VAR exhibits a lower average mean squared error with the target acoustic response and peak frequency variance compared to the parametrized inverse-designed VAR. Given that VAR is often required to attenuate hybrid noise with multi-peak frequency for practical application, we designed a multi-cavity VAR capable of broadband sound attenuation with multi-target peak frequency by combining the non-parametrized inverse-designed VARs.

## 2. Structural Design and Simulation of VAR

### 2.1 Acoustic Structure

Figure 1 shows the structures of the proposed axisymmetric VAR that attenuates the sound passing through it. The peak frequency of sound attenuation can be tuned by manipulating the geometric shape of the VAR's cavity. Figure 1a shows the schematic of a parameterized VAR which is defined by geometric parameters. The training dataset for the inverse design model proposed in this study consists of these parameterized VAR. As a result of our proposed inverse design model, a non-parameterized VAR is created as shown in Figure 1b. Non-parameterized VAR has a similar shape to parameterized VAR but has a non-typical structure that can not be defined with geometric parameters. Figures 1c and d show the acoustic domain of the unit VAR and the cross-section image with the unit VAR's geometric parameters, respectively. The geometric parameters of the VAR were bound to keep the typical T-shape cross-section image represented in Figure 1. The length of the air-fluid channel was fixed at $l_c$ = 20 mm. The range of other design parameters is listed in Table 1. $R, l_b, R_n, l_a, R_c$ represents the radius of the air-fluid channel, neck width of cavity's cross-section, height of cavity neck's cross-section, the width of the cavity, the total radius of VAR's acoustic domain respectively. Broadband sound attenuation with multi-peak frequency can be achieved by serially combining multiple unit VARs with different peak frequencies as shown in Figure 1e.

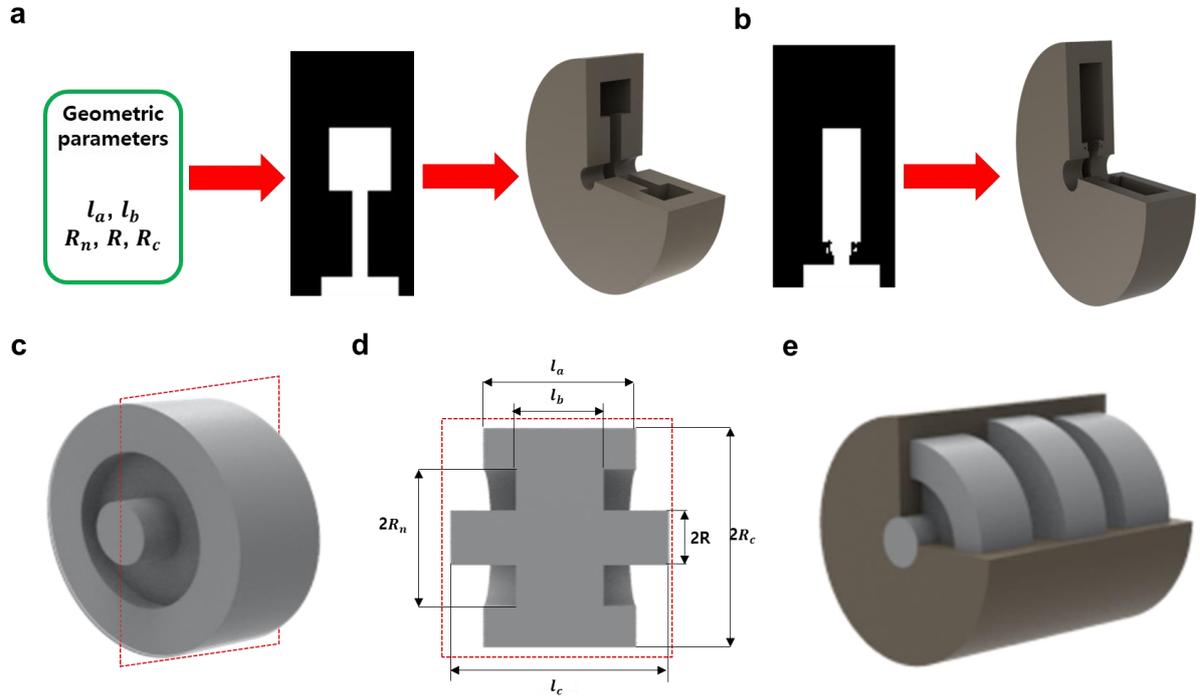

**Fig. 1** (a) Schematic of parameterized VAR (b) Schematic of non-parameterized VAR (c) Acoustic domain of the unit VAR (d) Cross-section image of the acoustic domain and design parameters of the unit VAR (e) Structure of a multi-cavity VAR

## 2.2 Numerical Simulation for Acoustic Response

We conducted finite element simulation using the frequency domain study in COMSOL Multiphysics 6.0 to obtain the acoustic response of the proposed VAR. The acoustic pressure field $p$ of the proposed VAR follows the Helmholtz equation below [58].

$$\nabla^2 p - \frac{1}{c^2}\frac{\partial^2 p}{\partial t^2} = 0 \qquad (1)$$

where $\nabla^2 = \frac{\partial^2}{\partial r^2} + \frac{1}{r}\frac{\partial}{\partial r} + \frac{1}{r^2}\frac{\partial^2}{\partial \theta^2} + \frac{\partial^2}{\partial z^2}$ is Laplacian operator in cylindrical coordinates, $c$(=343 m/s) is the speed of sound in air, and $\frac{\partial^2}{\partial t^2}$ is the second derivative with respect to time $t$. This equation is computed with 2D-axisymmetric as shown in Figure 2. The boundary of the VAR is assumed to be a sound-hard wall in consideration of the large impedance mismatch between the structure and air. Non-reflecting boundary conditions are applied to both ends of the waveguide. The plane wave, denoted as $p_i = P_i e^{j(\omega t - kz)}$ with the pressure amplitude $P_i$ (=1 Pa), a wavenumber $k$, and an angular frequency $\omega$ is applied to the inlet domain for the input wave. For simplicity, the loss of the VAR model is not considered. The attenuation performance of the VAR was evaluated based on the acoustic power of the inlet and outlet by numerically solving the governing equation. The sound transmission loss (STL) was used as an index to evaluate the soundproofing performance of the VAR expressed as

$$\text{STL} = 10 \log \frac{W_{in}}{W_{out}} \ [dB] \qquad (2)$$

where $W_{in}$ ($=\frac{|p^2_{@\,inlet}|}{2\rho c}$) and $W_{out}$ ($=\frac{|p^2_{@\,outlet}|}{2\rho c}$) represent the acoustic powers at the inlet and outlet, respectively, obtained using the governing equations, and $\rho$ is the density of air. 50 STL data points of each VAR were obtained by conducting simulations at 1-1961Hz and 40 Hz intervals. (see Supporting Information 1 for more detailed information about analytical solutions and validation of numerical simulation)

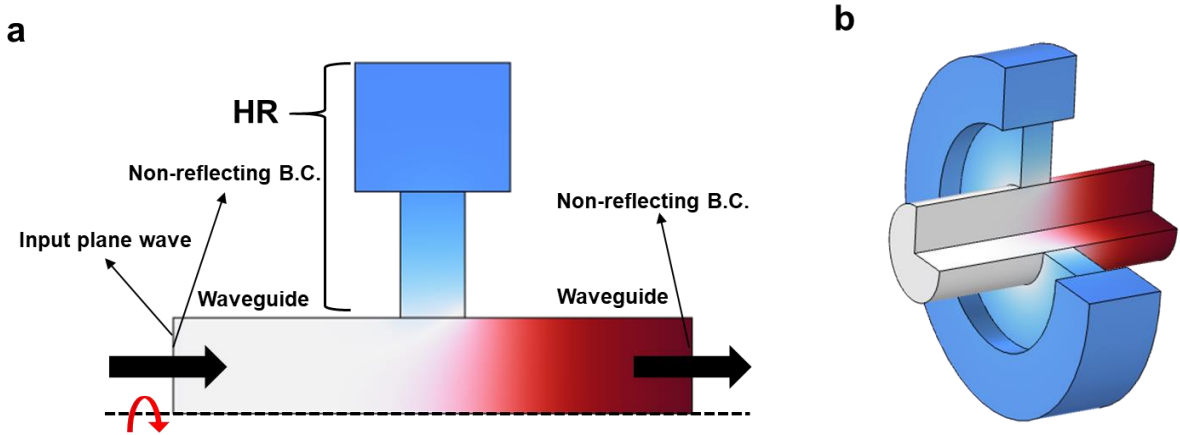

**Fig. 2** (a) Cross-section image of VAR numerical simulation(air domain) (b) 3D structure of VAR numerical simulation

## 3. Deep Generative Model for VAR Inverse-design

Many acoustic structures exhibit the same acoustic response; hence, the forward prediction of the acoustic response of the VAR should be considered a many-to-one mapping problem. However, the inverse-design problem is considered a one-to-many mapping problem because various geometric shapes could be designed for a given acoustic response. [59] To solve the one(acoustic response)-to-many(VAR structure) mapping problem of designing a VAR structure with the target acoustic response, we propose an inverse-design framework using a deep generative model as shown in Figure 3.

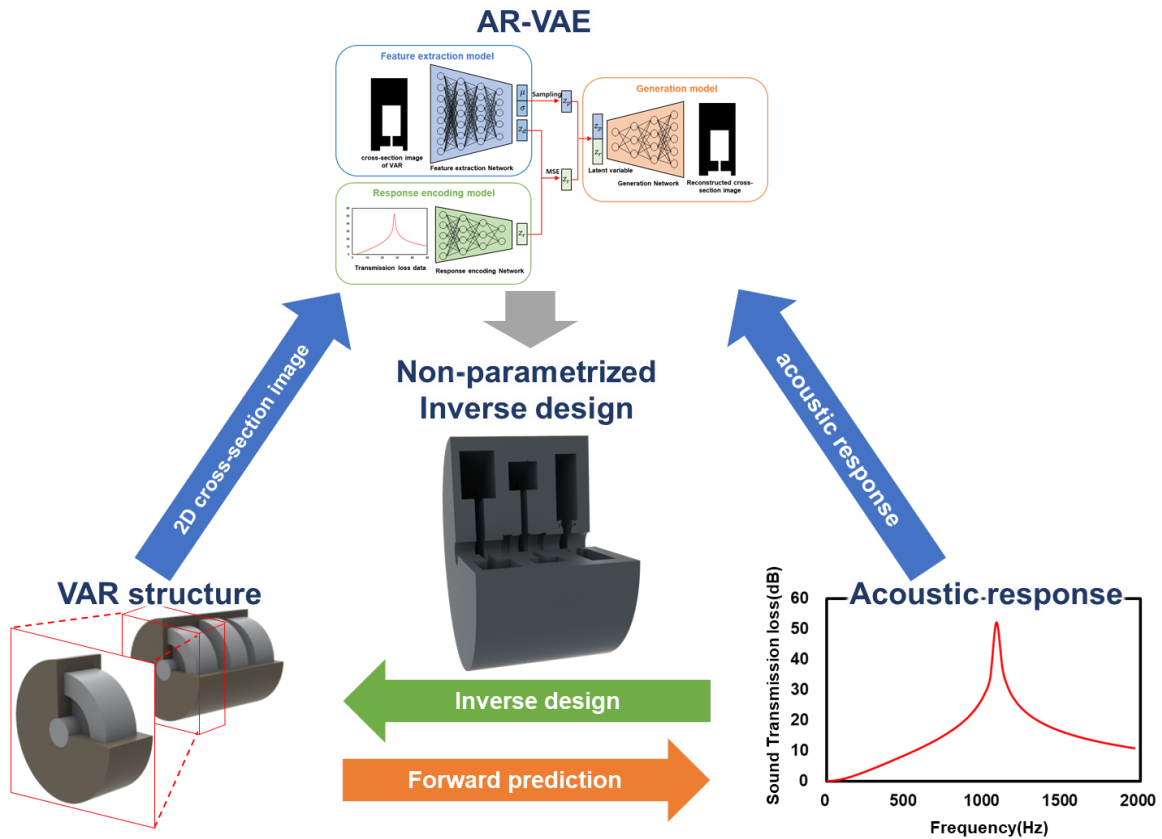

**Fig. 3** Schematic of the proposed inverse-design framework

## 3.1 Data Acquisition

We generated the dataset of the VAR cross-section images and the corresponding acoustic response data to train the AR-VAE. Since the proposed VAR has an axisymmetric shape, the cross-section image of the VAR structure was used as the input data for the deep generative model. The dataset of VAR cross-section images was randomly sampled within the design parameter range listed in Table 1. The cross-section images were expressed as $128 \times 64$ grayscale binary images, where 1 represents the acoustic domain inside the VAR and 0 represents the hard wall of the VAR. The corresponding acoustic response data, obtained by the numerical simulation, were converted into a $50 \times 1$ vector and used for training. A total of 53,350 data were generated with different design parameters and split into a training dataset (70% of the total data, 37,345) and a test dataset (30% of the total data, 16,005).

## 3.2 Acoustic Response Encoded VAE (AR-VAE)

The VAE is a deep generative model that consists of a bottleneck-shaped neural network with an encoder and a decoder. The VAE forms a latent space z, a dimension-reduced space, which encodes the probability distribution of the input data x to generate data similar to the input data x. When the VAE is sufficiently trained, the decoder can generate data similar to the input data x by inputting a variable sampled from the latent space z. To implement the data-generating feature of the VAE, training proceeds through the loss function below [54].

$$L(\emptyset, \theta; x) = -E_{q_\emptyset(z|x)}[\log\{p_\theta(x|z)\}] + D_{KL}(q_\emptyset(z|x) \| p_\theta(z)) \qquad (3)$$

where $\emptyset$ and $\theta$ represents the parameters of the encoder and decoder respectively. The loss function consists of a reconstruction term (the first term of the equation) and a regularization term (the second term of the equation). The reconstruction term indicates how well the input data x was reconstructed using the latent space z sampled by the encoder(approximate posterior) $q_\emptyset(z|x)$. The regularization term indicates the degree of similarity between the prior distribution $p_\theta(z)$ and approximate posterior $q_\emptyset(z|x)$. The regularization term prevents the VAE from simply performing a reconstruction task on the input data, by regularizing $\emptyset$ so that

the approximate posterior $q_\emptyset(z|x)$ can be close to the prior distribution $p_\theta(z)$. (see Supporting Information 2 for more detailed information on the VAE loss function)

Given that conventional VAEs do not contain acoustic response information, they can only generate VAR cross-section images of various shapes without any consideration of the acoustic response. To overcome the absence of acoustic response information, we introduced the novel VAE-based inverse-design model, AR-VAE. We added the acoustic response encoding model, which encodes the acoustic response to deterministic dimension-reduced latent space. Rather than simply matching the VAR cross-section image and acoustic response, it is more effective to extract and match important features between the VAR cross-section image and acoustic response through dimension reduction. [60, 61] In addition, to solve the one-to-many mapping problem mentioned above, it is necessary to match the probabilistic latent space that can generate various results with the acoustic response. Therefore, we trained AR-VAE to be able to match VAR's geometric information-encoded probabilistic latent space and VAR's acoustic response-encoded deterministic latent space. Consequently, the decoder was able to generate various VAR cross-section images with target acoustic response and also solved the one-to-many mapping problem.

Figure 4 shows the overall architecture of the AR-VAE, consisting of three deep neural network-based submodels— the feature extraction model, the response encoding model, and the generation model (see Supporting Information 3 for the detailed AR-VAE network architecture).

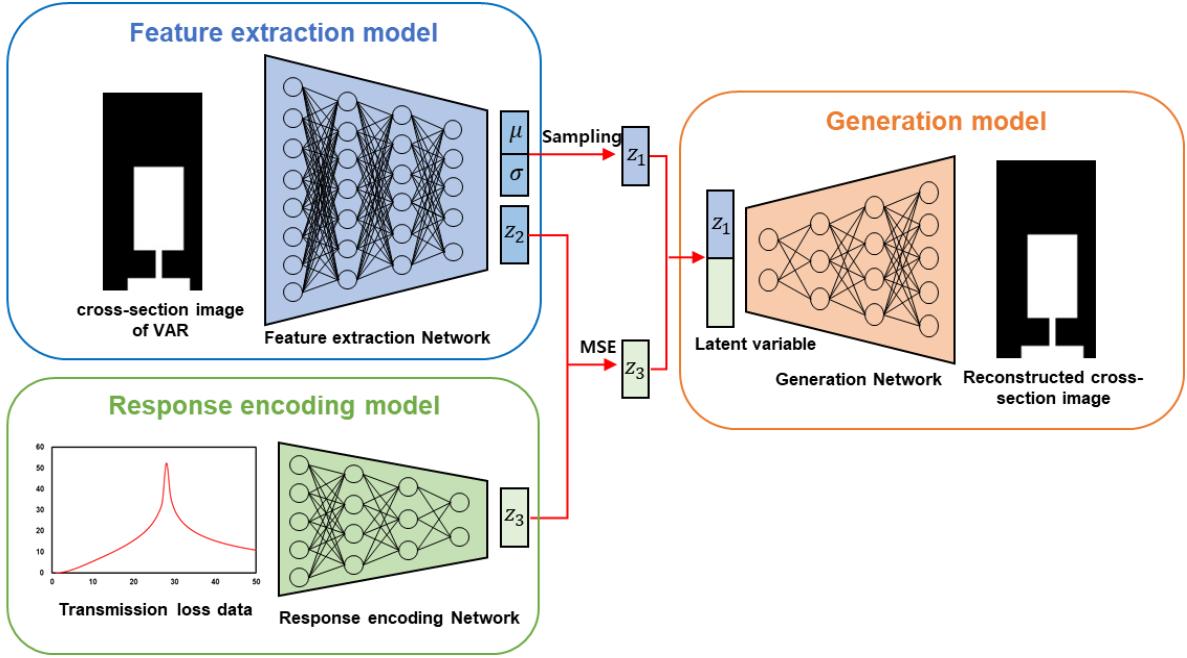

**Fig. 4** Overall architecture of the AR-VAE

*Feature extraction model*: The feature extraction model extracted the geometric features of the VAR cross-section image and encoded the features into a dimension-reduced latent space. A residual network, called ResNet, was introduced in the deep neural network of AR-VAE to prevent gradient vanishing or exploding problems during extracting features with relatively complex cross-sectional shapes [62]. The feature extraction network had a total of 24 output nodes, 16 of which were used to construct the probabilistic latent space $z_p$, which encodes the geometric feature information of the VAR cross-section image. The remaining 8 nodes were used to construct the deterministic latent space $z_d$, which was used to match the information between the geometric features and acoustic response.

*Response encoding model*: The response encoding model encodes the acoustic response data of the VAR into the dimension-reduced latent space $z_r$. The $50 \times 1$ acoustic response data obtained through the numerical simulation were used as the input. By extracting the features of the corresponding acoustic responses using a 1D convolution layer, an 8-dimensional deterministic latent space $z_r$ was constructed.

*Generation model*: The generation model uses both variables of the latent space $z_p$ and $z_r$ as inputs and aims to reconstruct a VAR cross-section image based on an input of a VAR cross-section image and a corresponding acoustic response. Reconstruction from a low-dimensional latent space to a high-dimensional image was performed based on the interpolation process of a transposed convolution layer.

The loss function of the conventional VAE was intended to reconstruct the VAR cross-section image without considering the acoustic response, and thus, the conventional VAE would be difficult to match the acoustic response and geometric feature of VAR. To resolve this issue, we devised a latent loss that is defined as the mean squared error (MSE) between the latent space $z_d$ constructed using the feature extraction model and the latent space $z_r$ constructed using the response encoding model and added the latent loss to the loss function of AR-VAE. Finally, the loss function of AR-VAE can be presented as follows.

$$L(\emptyset, \theta, \tau; x, y) = -E_{q_\emptyset(z|x)}[\log\{p_\theta(x|z_p, z_r)\}] + D_{KL}(q_\emptyset(z|x) \| p_\theta(z)) + \|f_\tau(y) - z_d\|^2 \quad (4)$$

where $z$ is the concatenation of $z_p$ and $z_d$; $z_r$ is a deterministic latent space variable encoded by the response encoding model $f_\tau$; $x$ and $y$ denote the VAR cross-section image data and acoustic response data, respectively. The response encoding model parameter $\tau$, feature extraction model parameter $\emptyset$, and generation model parameter $\theta$ were jointly learned by the backpropagation algorithm. By minimizing the above loss function, the AR-VAE could generate the VAR cross-section images reflecting the given acoustic response.

## 3.3 Training Process

The training process of AR-VAE proceeded by using an Intel® Core™ i5-8500 CPU @ 3.00 GHz and NVIDIA GeForce RTX3060 GPU. We set the batch size to 32, used Adam optimizer with a learning rate of 1e-5, and proceeded with training up to 12000 epochs. The training results are shown as the learning curves in Figure 5. The total loss is the sum of the reconstruction loss, the regularization loss, and the latent loss.

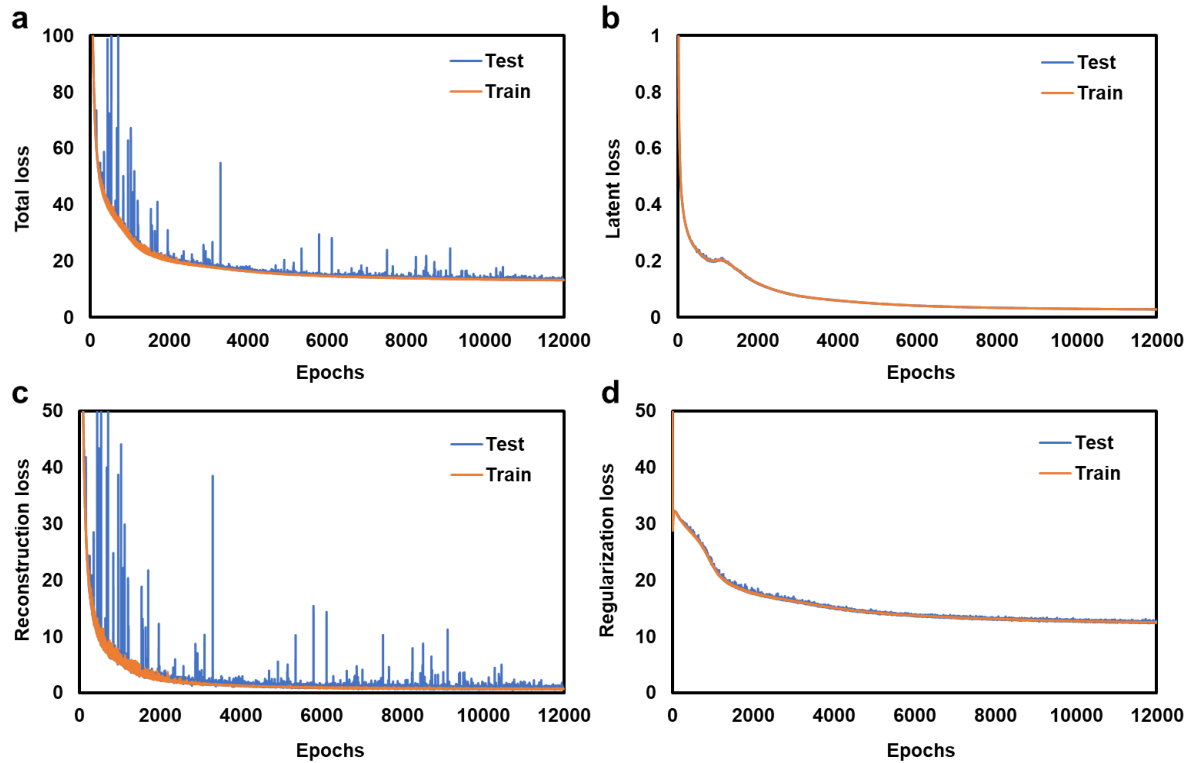

**Fig. 5** Learning curves of the AR-VAE (a) total loss, (b) latent loss, (c) reconstruction loss, and (d) regularization loss

## 4. Results and Discussion

### 4.1 Validation of AR-VAE

A validation process was conducted to assess the ability of the trained generation model to inverse-design a VAR cross-section image with an acoustic response similar to the input target acoustic response.

#### 4.1.1 Inverse-designed VAR by AR-VAE

The generation model of the AR-VAE created VAR cross-section images by inputting variables of $8 \times 1$ probabilistic latent space $z_p$ and $8 \times 1$ deterministic latent space $z_r$. To confirm the creation of various VAR cross-section images, we input variables of probabilistic latent space $z_p$ randomly sampled from a uniform distribution between -1 and 1 to the generation model with the target acoustic response. By inputting 100 randomly sampled $z_p$ for each given target acoustic response, 100 VAR candidates with acoustic responses similar to the target acoustic response were generated. Because the images generated by the AR-VAE are blurry, the generated image was converted to a binary image by thresholding. The acoustic response of the generated VAR candidates was evaluated based on the numerical simulation. Among 100 generated VAR candidates, the VAR with the smallest mean squared error (MSE) between the corresponding and target acoustic response was selected as an inverse-design result. We found that 100 sampling of VAR candidates could achieve a sufficiently smaller mean squared error between the corresponding and target acoustic response compared to that of training datasets. Figure 6 shows the schematic of the VAR inverse-design process using the AR-VAE. We confirmed that the corresponding acoustic response of generated VAR candidates was positioned close to the target acoustic response. (see Supporting Information 4 for more detailed information on the non-parameterized VAR inverse-design process by AR-VAE)

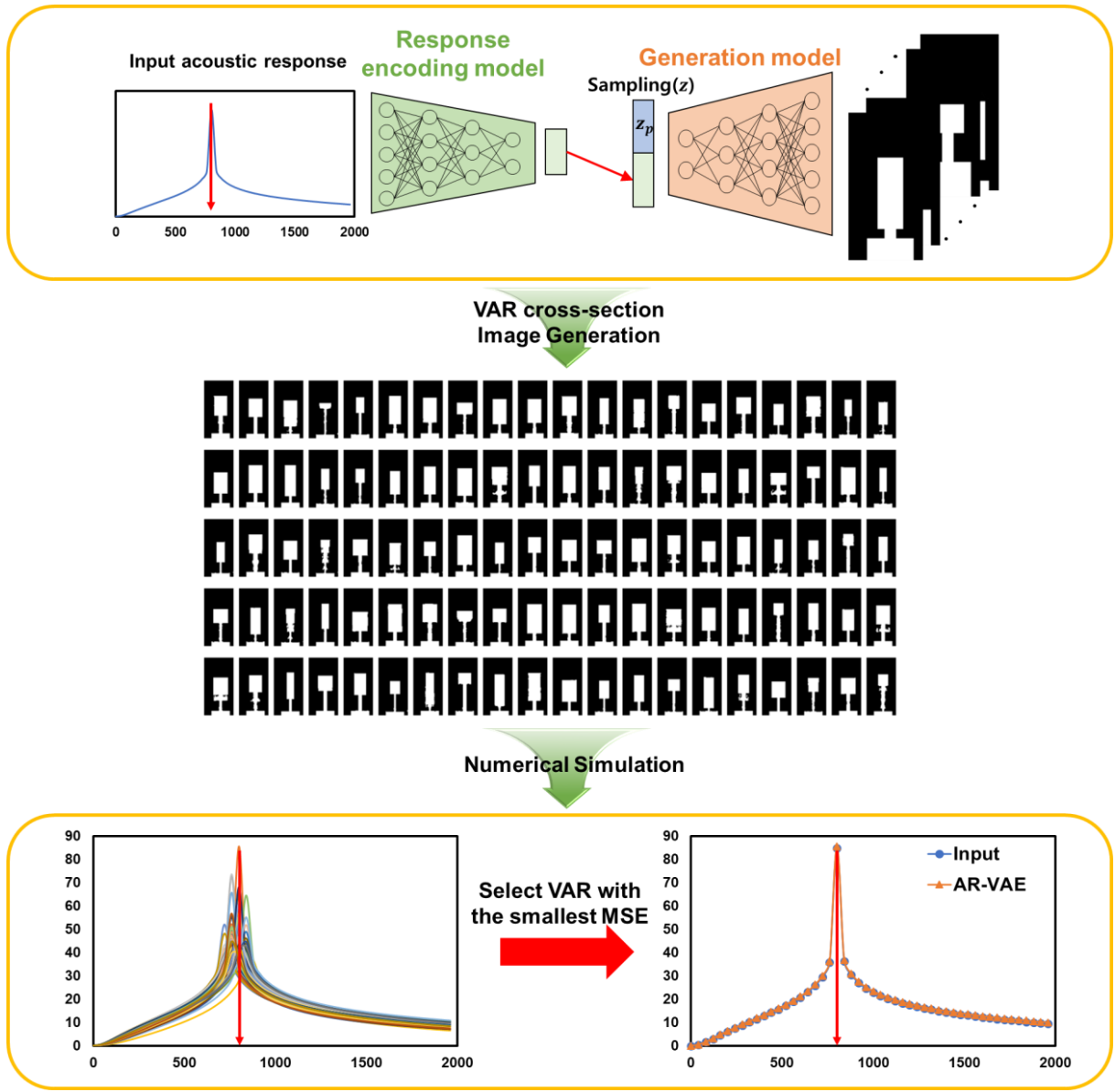

**Fig. 6** Schematic of the VAR inverse-design process using the AR-VAE

### 4.1.2 Validation of AR-VAE's inverse design performance

To verify the performance of AR-VAE, the target acoustic responses with peak frequencies of 601, 801, 1001, 1201, 1401, and 1601 Hz were selected from the test dataset. We inverse-designed VAR cross-section images corresponding to the selected target acoustic response by applying the inverse design method described above. We trained an acoustic response-parameter matching neural network (APNN ), which consisted of a simple deep neural network for inverse designing VAR by deterministically matching the VAR's geometric parameters and acoustic response as a control group. Because most of the previous research on AMs inverse design was based on deep learning-based parameter searching like APNN, it is considered appropriate to use APNN as a control group for AR-VAE performance validation. (see Supporting Information 5 for more detailed information on the APNN) [25, 55-57, 63] We also compared inverse-designed VAR by AR-VAE with the best VAR candidate in the training dataset, which has the smallest MSE with the target acoustic response in the training dataset to confirm that AR-VAE has better inverse design performance than the training dataset. The acoustic response of the inverse-designed VAR by AR-VAE, the inverse-designed VAR by APNN, and the best VAR candidate in the training dataset were compared, as shown in Figure 7. Figure 8 exhibited sound transmission loss error (STL error), an acoustic response difference with target acoustic response for each frequency, of the inverse-designed VAR by the AR-VAE, the inverse-designed VAR by the APNN, and the best VAR candidates in the training dataset. For all six target acoustic responses, the inverse-designed VAR by the AR-VAE exhibited the lowest STL error over the entire frequency range, particularly at the peak frequency. To compare the results quantitatively, the mean squared STL error is shown in Table 2. and it can be found that AR-VAE has the smallest mean squared STL error for all target acoustic responses. The inverse-designed VARs by the AR-VAE exhibit a 25-fold reduction in mean squared STL error compared to the APNN on average and a 2.5-fold reduction in mean squared STL error compared to the acoustic response of the best VAR candidate in the training dataset on average. Since AR-VAE performs inverse design based on binary cross-section images rather than the geometric parameters of VAR, various cross-section images with non-typical shapes are created as shown in Figure 6. To verify that inverse design using these non-parameterized VAR cross-section images can achieve a more accurate inverse design than using geometric parameters of VAR, we compared non-parameterized inverse-designed VAR with parameterized inverse-designed VAR which parameterizes non-parameterized inverse-designed VAR by detecting parameters from its cross-section images. (see Supporting Information 6 for more detailed information and results of the parameterized VAR cross-section image) The acoustic response of 100 non-parameterized inverse-designed VAR and 100 parameterized inverse-designed VAR for each target acoustic response were obtained by the numerical simulation. It can be confirmed that non-parameterized inverse-designed VAR has a lower average mean squared STL error for all six target acoustic responses. To confirm the reliability of peak frequency coincidence with the target acoustic response, the peak frequency variance of 100 acoustic responses of non-parameterized inverse-designed VAR and parameterized inverse-designed VAR were compared. It was confirmed that the non-parameterized inverse-designed VAR has a lower peak frequency variance, which confirms that the non-parameterized inverse-designed VAR can produce more reliable inverse-design results. (see Supporting Information 7 for average mean squared STL error and peak frequency variance comparison results between non-parameterized inverse-designed VAR and parameterized inverse-designed VAR)

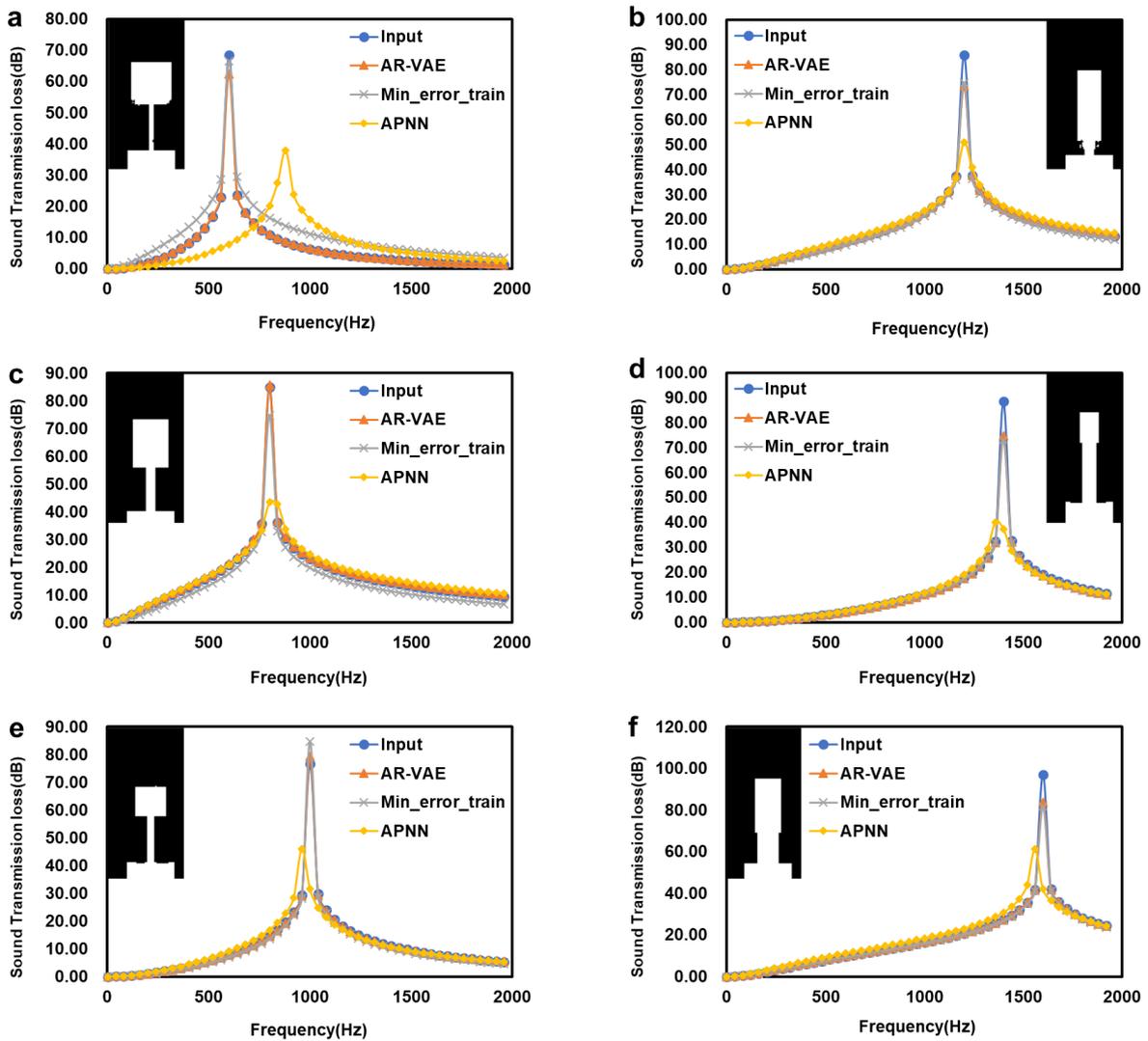

**Fig. 7** Comparison between the target acoustic response (blue, circle), the acoustic response of inverse designed VAR by AR-VAE (orange, triangle), the best VAR candidate in training dataset (gray, cross mark), and the acoustic response of inverse designed VAR by APNN (yellow, rhombus) at peak frequencies (a) 601 Hz, (b) 1201 Hz, (c) 801 Hz, (d) 1401 Hz, (e) 1001 Hz, and (f) 1601 Hz

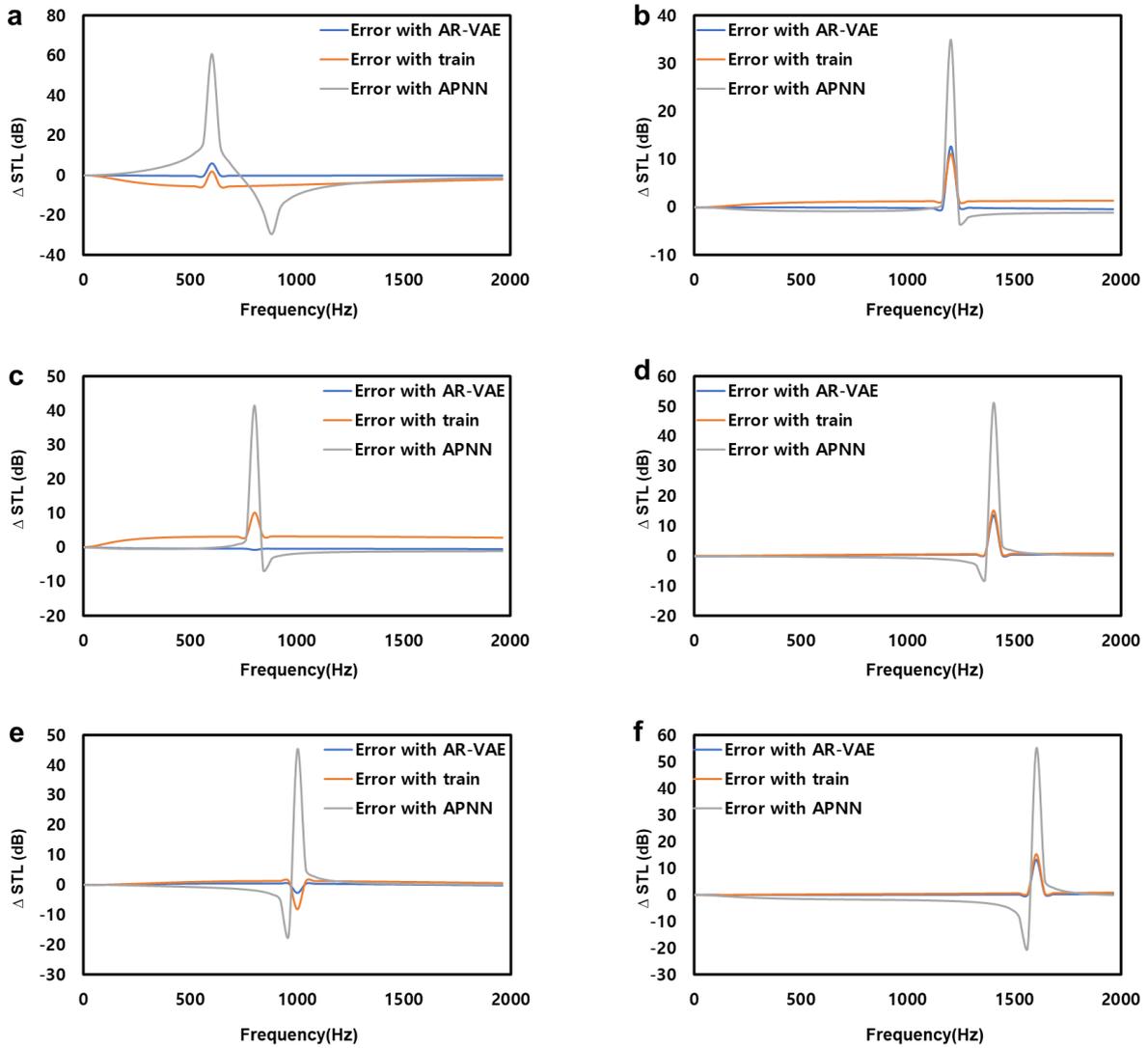

**Fig. 8** Graphs representing the STL error of the inverse-designed VAR by AR-VAE (blue), the best VAR candidate in training dataset (orange), and inverse-designed VAR by APNN (gray) for each target acoustic response at peak frequencies (a) 601 Hz, (b) 1201 Hz, (c) 801 Hz, (d) 1401 Hz, (e) 1001 Hz, and (f) 1601 Hz

## 4.2 Multi-cavity VAR

Due to the hybrid noise in environments where both noise reduction and ventilation are required, broadband sound attenuation with multi-peak frequency is often required. To realize the broadband sound attenuation performance with multi-peak frequency, a serial array of resonators should be implemented. The interaction between each single resonator composing a serial resonator array can be analyzed by the transfer matrix method. According to this principle, broadband sound attenuation is realized while preserving the peak frequency of each single resonator. [64-68] A multi-cavity VAR was created by integrating multiple non-parameterized inverse-designed unit VARs. The structure and acoustic response of the multi-cavity VAR are presented in Figure 9. The numerical simulation revealed that the peak frequencies of the individual unit VARs were maintained in the multi-cavity VAR. Furthermore, a broadband attenuation performance within the range of 601 Hz – 1961 Hz was confirmed.

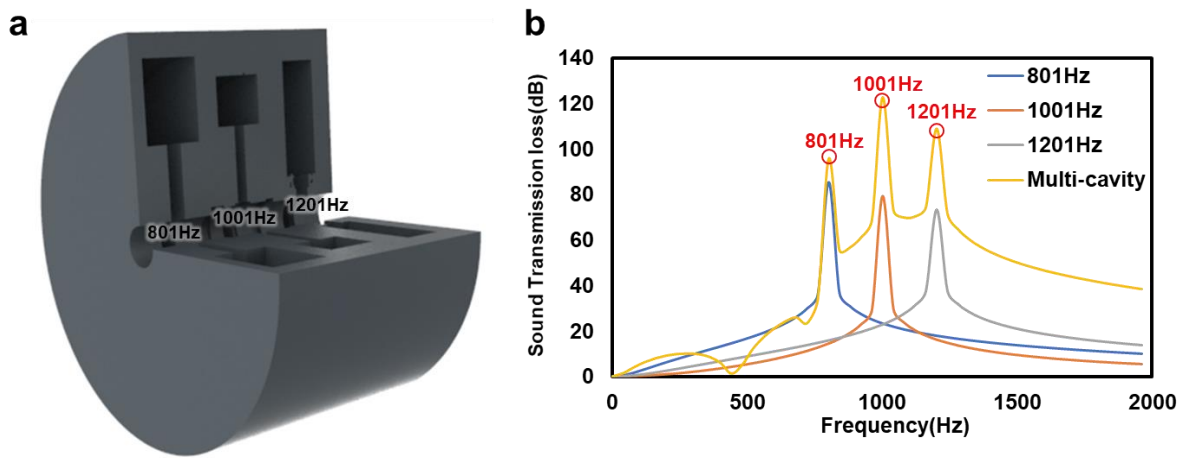

**Fig. 9** Structure and acoustic response of inverse-designed multi-cavity VAR

## 5. Conclusion

This study proposes a novel deep generative model-based method for the inverse design of non-parameterized VARs. We proposed a VAE-based inverse-design model called the AR-VAE for accurate matching between VAR's geometric shapes and acoustic response by realizing the non-parameterized design of VAR. By combining the inverse-designed non-parameterized VARs, a multi-cavity VAR structure capable of broadband and multi-peak frequency attenuation was realized. Therefore, the method proposed in this study enables the accurate inverse design of a VAR structure that attenuates multi-target peak frequencies. Furthermore, the generative characteristics of the AR-VAE can be exploited to design structures suitable for other physical constraints and non-parametric designs. Thus, the AR-VAE can potentially be applied to the inverse design of structures that exhibit high-dimensional nonlinear physical responses, beyond acoustic response.

## Competing Interests

The authors declare that they have no known competing financial interests or personal relationships that could have appeared to influence the work reported in this paper.

## Acknowledgments

This study has been conducted with the support of the Ministry of Trade, Industry, and Energy (MOTIE) of Korea through the "Innovative Digital Manufacturing Platform" (project no. P0022331) supervised by the Korea Institute for Advancement of Technology (KIAT), and the Korea Institute of Energy Technology Evaluation and Planning (KETEP) (No.20227310100010). This work was supported by the National Research Foundation of Korea (NRF) grant funded by the Korea government (MSIT) (No. RS-2023-00209094).

# Tables

**Table 1.** Geometric parameters of VAR.

|  | Min | Max |
|---|---|---|
| $l_a$ | 4 | 18 |
| $l_b$ | 2 | $l_a$-1 |
| $R$ | 5 | 20 |
| $R_C$ | $\frac{R+6}{2}$ | 48.5 |
| $R_n$ | 7 | $\frac{R_c - R + 38}{2}$ |

**Table 2.** Mean squared STL error comparison for each target acoustic response

| Peak frequency | The best VAR candidate in the training dataset | Inverse-designed VAR by AR-VAE | Inverse-designed VAR by APNN |
|---|---|---|---|
| 601Hz | 16.477 | 0.966 | 131.981 |
| 801Hz | 9.882 | 0.205 | 36.583 |
| 1001Hz | 2.179 | 0.224 | 48.224 |
| 1201Hz | 3.862 | 3.236 | 25.798 |
| 1401Hz | 4.958 | 3.898 | 54.401 |
| 1601Hz | 4.995 | 3.52 | 73.589 |

# References

[1] Munjal ML,(1998), Analysis and Design of Mufflers—An Overview of Research at The Indian Institute of Science. *journal of Sound and Vibration,* vol. 211, no. 3, pp. 425-433, Art no., doi: 10.1006/jsvi.1997.1309

[2] Munjal ML,(2003), Analysis and design of pod silencers. *Journal of Sound and Vibration,* vol. 262, no. 3, pp. 497-507, Art no., doi: 10.1016/s0022-460x(03)00108-1

[3] Yu X and L Cheng,(2015), Duct noise attenuation using reactive silencer with various internal configurations. *Journal of Sound and Vibration,* vol. 335, pp. 229-244, Art no., doi: 10.1016/j.jsv.2014.08.035

[4] Guérin S, E Thomy, and MCM Wright,(2005), Aeroacoustics of automotive vents. *Journal of Sound and Vibration,* vol. 285, no. 4-5, pp. 859-875, Art no., doi: 10.1016/j.jsv.2004.08.043

[5] Cho Y, S Wang, J Hyun, S Oh, and S Goo,(2018), Analysis of sound absorption performance of an electroacoustic absorber using a vented enclosure. *Journal of Sound and Vibration,* vol. 417, pp. 110-131, Art no., doi: 10.1016/j.jsv.2017.11.051

[6] Liu E, S Peng, and T Yang,(2018), Noise-silencing technology for upright venting pipe jet noise. *Advances in Mechanical Engineering,* vol. 10, no. 8, doi: 10.1177/1687814018794819

[7] Kang J,(2006), An Acoustic Window System with Optimum Ventilation and Daylighting Performance. *Noise & Vibration Worldwide,* vol. 37, no. 11, pp. 9-17, Art no., doi: 10.1260/095745606779385108

[8] Huang H, X Qiu, and J Kang,(Jul 2011), Active noise attenuation in ventilation windows. *J Acoust Soc Am,* vol. 130, no. 1, pp. 176-88, Art no., doi: 10.1121/1.3596457

[9] Wang X, X Luo, B Yang, and Z Huang,(2019), Ultrathin and durable open metamaterials for simultaneous ventilation and sound reduction. *Applied Physics Letters,* vol. 115, no. 17, doi: 10.1063/1.5121366

[10] Wang S, J Tao, X Qiu, and IS Burnett,(Mar 18 2021), Broadband noise insulation of windows using coiled-up silencers consisting of coupled tubes. *Sci Rep,* vol. 11, no. 1, p. 6292, Art no., doi: 10.1038/s41598-021-85796-0

[11] Yang Z, J Mei, M Yang, NH Chan, and P Sheng,(Nov 14 2008), Membrane-type acoustic metamaterial with negative dynamic mass. *Phys Rev Lett,* vol. 101, no. 20, p. 204301, Art no., doi: 10.1103/PhysRevLett.101.204301

[12] Huang HH, CT Sun, and GL Huang,(2009), On the negative effective mass density in acoustic metamaterials. *International Journal of Engineering Science,* vol. 47, no. 4, pp. 610-617, Art no., doi: 10.1016/j.ijengsci.2008.12.007

[13] Lee SH, CM Park, YM Seo, ZG Wang, and CK Kim,(2009), Acoustic metamaterial with negative density. *Physics Letters A,* vol. 373, no. 48, pp. 4464-4469, Art no., doi: 10.1016/j.physleta.2009.10.013

[14] Fang N *et al.*,(Jun 2006), Ultrasonic metamaterials with negative modulus. *Nat Mater,* vol. 5, no. 6, pp. 452-6, Art no., doi: 10.1038/nmat1644

[15] Akl W and A Baz,(2010), Multi-cell Active Acoustic Metamaterial with Programmable Bulk Modulus. *Journal of Intelligent Material Systems and Structures,* vol. 21, no. 5, pp. 541-556, Art no., doi: 10.1177/1045389x09359434

[16] García-Chocano VM, R Graciá-Salgado, D Torrent, F Cervera, and J Sánchez-Dehesa,(2012), Quasi-two-dimensional acoustic metamaterial with negative bulk modulus. *Physical Review B,* vol. 85, no. 18, doi: 10.1103/PhysRevB.85.184102

[17] Kim S-H and S-H Lee,(2014), Air transparent soundproof window. *AIP Advances,* vol. 4, no. 11, doi: 10.1063/1.4902155

[18] Ge Y, H-x Sun, S-q Yuan, and Y Lai,(2019), Switchable omnidirectional acoustic insulation through open window structures with ultrathin metasurfaces. *Physical Review Materials,* vol. 3, no. 6, doi: 10.1103/PhysRevMaterials.3.065203

[19] Lee T, T Nomura, EM Dede, and H Iizuka,(2019), Ultrasparse Acoustic Absorbers Enabling Fluid Flow and Visible-Light Controls. *Physical Review Applied,* vol. 11, no. 2, doi: 10.1103/PhysRevApplied.11.024022

[20] Fusaro G, X Yu, J Kang, and F Cui,(2020), Development of metacage for noise control and natural ventilation in a window system. *Applied Acoustics,* vol. 170, doi: 10.1016/j.apacoust.2020.107510

[21] Kumar S, TB Xiang, and HP Lee,(2020), Ventilated acoustic metamaterial window panels for simultaneous noise shielding and air circulation. *Applied Acoustics,* vol. 159, doi: 10.1016/j.apacoust.2019.107088

[22] Huang S, X Fang, X Wang, B Assouar, Q Cheng, and Y Li,(Jan 2019), Acoustic perfect absorbers via Helmholtz resonators with embedded apertures. *J Acoust Soc Am,* vol. 145, no. 1, p. 254, Art no., doi: 10.1121/1.5087128

[23] Molesky S, Z Lin, AY Piggott, W Jin, J Vucković, and AW Rodriguez,(2018), Inverse design in nanophotonics. *Nature Photonics,* vol. 12, no. 11, pp. 659-670, Art no., doi: 10.1038/s41566-018-0246-9

[24] Du X, J Ren, and L Leifsson,(2019), Aerodynamic inverse design using multifidelity models and manifold mapping. *Aerospace Science and Technology,* vol. 85, pp. 371-385, Art no., doi: 10.1016/j.ast.2018.12.008

[25] Gao N, M Wang, B Cheng, and H Hou,(2021), Inverse design and experimental verification of an acoustic sink based on machine learning. *Applied Acoustics,* vol. 180, doi: 10.1016/j.apacoust.2021.108153

[26] Jung JW, JE Kim, and JW Lee,(2018), Acoustic metamaterial panel for both fluid passage and broadband soundproofing in the audible frequency range. *Applied Physics Letters,* vol. 112, no. 4, doi: 10.1063/1.5004605

[27] Dell A, A Krynkin, and KV Horoshenkov,(2021), The use of the transfer matrix method to predict the effective fluid properties of acoustical systems. *Applied Acoustics,* vol. 182, doi: 10.1016/j.apacoust.2021.108259

[28] Håkansson A, J Sánchez-Dehesa, and L Sanchis,(2004), Acoustic lens design by genetic

algorithms. *Physical Review B,* vol. 70, no. 21, doi: 10.1103/PhysRevB.70.214302

[29] Li D, L Zigoneanu, BI Popa, and SA Cummer,(Oct 2012), Design of an acoustic metamaterial lens using genetic algorithms. *J Acoust Soc Am,* vol. 132, no. 4, pp. 2823-33, Art no., doi: 10.1121/1.4744942

[30] Krishna A, SR Craig, C Shi, and VR Joseph,(2022), Inverse design of acoustic metasurfaces using space-filling points. *Applied Physics Letters,* vol. 121, no. 7, doi: 10.1063/5.0096869

[31] Kojima K, B Wang, U Kamilov, T Koike-Akino, and K Parsons (2017) "Acceleration of FDTD-based inverse design using a neural network approach," in *Integrated Photonics Research, Silicon and Nanophotonics*: Optica Publishing Group, p. ITu1A. 4.

[32] Chen Y, F Meng, and X Huang,(2021), Creating acoustic topological insulators through topology optimization. *Mechanical Systems and Signal Processing,* vol. 146, doi: 10.1016/j.ymssp.2020.107054

[33] Chai J, H Zeng, A Li, and EWT Ngai,(2021), Deep learning in computer vision: A critical review of emerging techniques and application scenarios. *Machine Learning with Applications,* vol. 6, doi: 10.1016/j.mlwa.2021.100134

[34] Kim MJ, JY Song, SH Hwang, DY Park, and SM Park,(Sep 29 2022), Electrospray mode discrimination with current signal using deep convolutional neural network and class activation map. *Sci Rep,* vol. 12, no. 1, p. 16281, Art no., doi: 10.1038/s41598-022-20352-y

[35] Young T, D Hazarika, S Poria, and E Cambria,(2018), Recent Trends in Deep Learning Based Natural Language Processing [Review Article]. *IEEE Computational Intelligence Magazine,* vol. 13, no. 3, pp. 55-75, Art no., doi: 10.1109/mci.2018.2840738

[36] Han B-Z, W-X Huang, and C-X Xu,(2022), Deep reinforcement learning for active control of flow over a circular cylinder with rotational oscillations. *International Journal of Heat and Fluid Flow,* vol. 96, p. 109008, Art no.

[37] Hwang SH *et al.*,(2023), Adaptive Electrospinning System Based on Reinforcement Learning for Uniform-Thickness Nanofiber Air Filters. *Advanced Fiber Materials,* vol. 5, no. 2, pp. 617-631, Art no., doi: 10.1007/s42765-022-00247-3

[38] Kim Y and SH Park,(2023), Highly Productive 3D Printing Process to Transcend Intractability in Materials and Geometries via Interactive Machine-Learning-Based Technique. *Advanced Intelligent Systems*, doi: 10.1002/aisy.202200462

[39] Zhang Z, J Geiger, J Pohjalainen, AE-D Mousa, W Jin, and B Schuller,(2018), Deep learning for environmentally robust speech recognition: An overview of recent developments. *ACM Transactions on Intelligent Systems and Technology (TIST),* vol. 9, no. 5, pp. 1-28, Art no.

[40] Liu Z, D Zhu, SP Rodrigues, KT Lee, and W Cai,(Oct 10 2018), Generative Model for the Inverse Design of Metasurfaces. *Nano Lett,* vol. 18, no. 10, pp. 6570-6576, Art no., doi: 10.1021/acs.nanolett.8b03171

[41] Ma W, F Cheng, Y Xu, Q Wen, and Y Liu,(Aug 2019), Probabilistic Representation and Inverse Design of Metamaterials Based on a Deep Generative Model with Semi-Supervised Learning Strategy. *Adv Mater,* vol. 31, no. 35, p. e1901111, Art no., doi: 10.1002/adma.201901111

[42] Wang Z, W Xian, MR Baccouche, H Lanzerath, Y Li, and H Xu,(2022), Design of Phononic Bandgap Metamaterials Based on Gaussian Mixture Beta Variational Autoencoder and Iterative Model Updating. *Journal of Mechanical Design,* vol. 144, no. 4, doi: 10.1115/1.4053814

[43] Wang L, Y-C Chan, F Ahmed, Z Liu, P Zhu, and W Chen,(2020), Deep generative modeling for mechanistic-based learning and design of metamaterial systems. *Computer Methods in Applied Mechanics and Engineering,* vol. 372, doi: 10.1016/j.cma.2020.113377

[44] Felsch G, N Ghavidelnia, D Schwarz, and V Slesarenko,(2023), Controlling auxeticity in curved-beam metamaterials via a deep generative model. *Computer Methods in Applied Mechanics and Engineering,* vol. 410, p. 116032, Art no.

[45] Xue T, TJ Wallin, Y Menguc, S Adriaenssens, and M Chiaramonte,(2020), Machine learning generative models for automatic design of multi-material 3D printed composite solids. *Extreme Mechanics Letters,* vol. 41, doi: 10.1016/j.eml.2020.100992

[46] On H-I, L Jeong, M Jung, D-J Kang, J-H Park, and H-J Lee,(2021), Optimal design of microwave absorber using novel variational autoencoder from a latent space search strategy. *Materials & Design,* vol. 212, doi: 10.1016/j.matdes.2021.110266

[47] Lee S *et al.*,(2022), Machine learning-enabled development of high performance gradient-index phononic crystals for energy focusing and harvesting. *Nano Energy,* vol. 103, doi: 10.1016/j.nanoen.2022.107846

[48] Mahesh K, SK Ranjith, and R Mini,(2023), A deep autoencoder based approach for the inverse design of an acoustic-absorber. *Engineering with Computers,* pp. 1-22, Art no.

[49] Wu R-T, M Jokar, MR Jahanshahi, and F Semperlotti,(2022), A physics-constrained deep learning based approach for acoustic inverse scattering problems. *Mechanical Systems and Signal Processing,* vol. 164, doi: 10.1016/j.ymssp.2021.108190

[50] Donda K, Y Zhu, A Merkel, S Wan, and B Assouar,(2022), Deep learning approach for designing acoustic absorbing metasurfaces with high degrees of freedom. *Extreme Mechanics Letters,* vol. 56, doi: 10.1016/j.eml.2022.101879

[51] Wu J, X Feng, X Cai, X Huang, and Q Zhou,(2022), A deep learning-based multi-fidelity optimization method for the design of acoustic metasurface. *Engineering with Computers,* pp. 1-19, Art no.

[52] Donda K *et al.*,(2021), Ultrathin acoustic absorbing metasurface based on deep learning approach. *Smart Materials and Structures,* vol. 30, no. 8, doi: 10.1088/1361-665X/ac0675

[53] Creswell A, T White, V Dumoulin, K Arulkumaran, B Sengupta, and AA Bharath,(2018), Generative adversarial networks: An overview. *IEEE signal processing magazine,* vol. 35, no. 1, pp. 53-65, Art no.

[54] Kingma DP and M Welling,(2013), Auto-encoding variational bayes. *arXiv preprint arXiv:1312.6114*.

[55] Xuecong Sun HJ, Yuzhen Yang, Han Zhao, Yafeng Bi, Zhaoyong Sun, Jun Yang,(2021), Acoustic Structure Inverse Design and Optimization Using Deep learning. *arXiv:2102.02063*.

[56] Liu TW, CT Chan, and RT Wu,(Feb 24 2023), Deep-Learning-Based Acoustic Metamaterial Design for Attenuating Structure-Borne Noise in Auditory Frequency Bands. *Materials (Basel),* vol. 16, no. 5, doi: 10.3390/ma16051879

[57] Mahesh K, S Kumar Ranjith, and RS Mini,(2021), Inverse design of a Helmholtz resonator based low-frequency acoustic absorber using deep neural network. *Journal of Applied Physics,* vol. 129, no. 17, doi: 10.1063/5.0046582

[58] MUNJAL ML *et al.* (2008) *Formulas of acoustics.* Springer Science & Business Media.

[59] Wiecha PR, A Arbouet, C Girard, and OL Muskens,(2021), Deep learning in nano-photonics: inverse design and beyond. *Photonics Research,* vol. 9, no. 5, doi: 10.1364/prj.415960

[60] Wan Z *et al.,*(Feb 2023), Old Photo Restoration via Deep Latent Space Translation. *IEEE Trans Pattern Anal Mach Intell,* vol. 45, no. 2, pp. 2071-2087, Art no., doi: 10.1109/TPAMI.2022.3163183

[61] Nakkwan Choi SL, Yongsik Lee, Seungjoon Yang (2023) "Restoration_of_Hand Drawn_Architectural Drawings Using Latent Space Mapping With Degradation Generator," in *Proceedings of the IEEE/CVF Conference on Computer Vision and Pattern Recognition (CVPR)*, pp. 14164-14172.

[62] He K, X Zhang, S Ren, and J Sun (2016) "Deep residual learning for image recognition," in *Proceedings of the IEEE conference on computer vision and pattern recognition*, pp. 770-778.

[63] Li R, Y Jiang, R Zhu, Y Zou, L Shen, and B Zheng,(Jul 14 2022), Design of ultra-thin underwater acoustic metasurface for broadband low-frequency diffuse reflection by deep neural networks. *Sci Rep,* vol. 12, no. 1, p. 12037, Art no., doi: 10.1038/s41598-022-16312-1

[64] Seo S-H, Y-H Kim, and K-J Kim,(2016), Transmission loss of a silencer using resonator arrays at high sound pressure level. *Journal of Mechanical Science and Technology,* vol. 30, no. 2, pp. 653-660, Art no., doi: 10.1007/s12206-016-0119-4

[65] Sang-Hyun Seo Y-HK,(2005), Silencer design by using array resonators for low-frequency band noise reduction. *J Acoust Soc Am,* vol. 118, doi: 10.1121/1.2036222

[66] Cai C and CM Mak,(2018), Acoustic performance of different Helmholtz resonator array configurations. *Applied Acoustics,* vol. 130, pp. 204-209, Art no., doi: 10.1016/j.apacoust.2017.09.026

[67] Wu D, N Zhang, CM Mak, and C Cai,(2019), Hybrid noise control using multiple Helmholtz resonator arrays. *Applied Acoustics,* vol. 143, pp. 31-37, Art no., doi: 10.1016/j.apacoust.2018.08.023

[68] Nguyen H, Q Wu, X Xu, H Chen, S Tracy, and G Huang,(2020), Broadband acoustic silencer with ventilation based on slit-type Helmholtz resonators. *Applied Physics Letters,* vol. 117, no. 13, doi: 10.1063/5.0024018

Supporting Information for

**Inverse design of Non-parameterized Ventilated Acoustic Resonator via Variational Autoencoder with Acoustic Response-encoded Latent Space**

Min Woo Cho[1,*], Seok Hyeon Hwang[1,*], Jun Young Jang[1,*], Jin Yeong Song[1], Sunkwang Hwang[2], Kyoung Je Cha[2], Dong Yong Park[2,**], Kyungjun Song[1,**], Sang Min Park[1,**]

[1]*School of Mechanical Engineering, Pusan National University, 63-2 Busan University-ro, Geumjeong-gu, Busan, 46241, South Korea*

[2]*Smart Manufacturing Technology R&D Group, Korea Institute of Industrial Technology, 320 Techno sunhwan-ro, Yuga-eup, Dalseong-gun, Daegu, 42994, South Korea*

[*]*Equally contributed*

[**]Corresponding author:

sangmin.park@pusan.ac.kr (Sang Min Park)

dypark9606@kitech.re.kr (Dong Yong Park)

song3396@pusan.ac.kr (Kyungjun Song)

1. Analytical validation of VAR numerical simulation

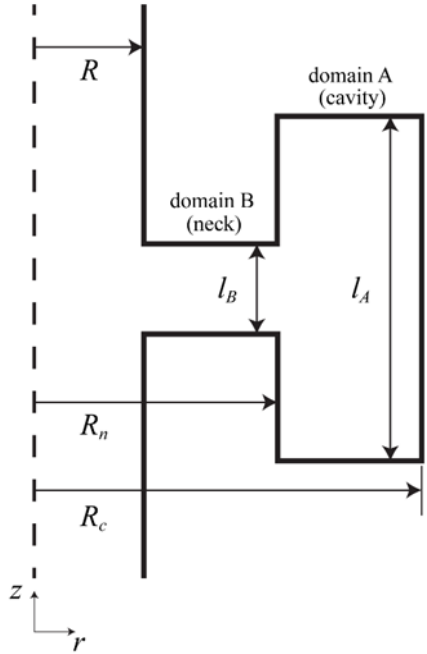

**Fig S1.** A simple HR model in cylindrical coordinates.

The following figure shows the HR model attached radially in a waveguide. HR is assumed to be excited only for the axisymmetric mode (plane wave radiation), and wave propagation in the z-direction is ignored inside the HR. In the cavity domain (domain A), the sound pressure and particle velocity are expressed as follows:

$$p_A = C_A[J_0(k_r r) - \alpha_A N_0(k_r r)] \tag{1a}$$

$$v_A = -j\frac{C_A}{\rho c}[J_1(k_r r) - \alpha_A N_1(k_r r)] \tag{1b}$$

where $k_r\ (=\sqrt{k^2 - k_z^2})$ is the radial wavenumber, $J_n$ and $N_n$ are the first and second kind Bessel function with order $n$, $j = \sqrt{-1}$, $\rho$ is the density of air, and $c$ is the speed of sound of air. At $r = R_c$, the particle velocity is zero (sound hard wall), we can calculate the $\alpha_A$:

$$\alpha_A = \frac{J_1(k_r R_c)}{N_1(k_r R_c)} \tag{2}$$

In the neck domain (domain B), pressure and particle velocity are expressed as follows:

$$p_B = C_B[J_0(k_r r) - \alpha_B N_0(k_r r)] \quad (3a)$$

$$v_B = -j\frac{C_B}{\rho c}[J_1(k_r r) - \alpha_B N_1(k_r r)] \quad (3b)$$

At $r = R_n$, the acoustic pressure and volume velocity are continuous with each other in the cavity and neck domain, which leads to the following relationship between the specific acoustic impedance in the two domains:

$$z_A|_{r=R_n} = z_B|_{r=R_n} \cdot \frac{S_A}{S_B} \quad (4)$$

where $z_A = p_A/v_A$ and $z_B = p_B/v_B$ are the specific acoustic impedance of the cavity and neck domain. $S_A/S_B$ is the ratio of the cross-sectional area of domain A and B, which is equal to the width ratio $l_A/l_B$.

Due to (E4), $\alpha_B$ is calculated as follows:

$$\alpha_B = \frac{j\rho c(\frac{S_a}{S_b})J_0(k_r R_n) - z_A|_{r=R_n}J_1(k_r R_n)}{j\rho c(\frac{S_a}{S_b})N_0(k_r R_n) - z_A|_{r=R_n}N_1(k_r R_n)} \quad (5)$$

and the specific acoustic impedance at the inlet of HR is expressed as follows:

$$z_B|_{r=R} = j\rho c \frac{J_0(k_r R) - \alpha_B N_0(k_r R)}{J_1(k_r R) - \alpha_B N_1(k_r R)} \quad (6)$$

By the calculated specific acoustic impedance, the transmission coefficient [1] is expressed as follows:

$$T = 1 - \frac{\delta}{\delta + z_B|_{r=R}} \quad (7)$$

where $\delta$ ($=\sigma\rho c/2$) is the resistance part corresponding to the far-field radiation of HR, and $\sigma$ ($=2\pi l_B/\pi R^2$) is the space opening ratio of the HR. With Eq.(E7), the sound transmission loss (STL) of the sound insulation material is calculated by:

$$STL = 10\ log \left|\frac{1}{T}\right|^2 \tag{8}$$

To verify the sound insulation performance of the HR design from VAE, the STLs of the theoretical and numerical models were compared. Figure 2 shows the calculated STLs of the two models when the design parameters $R_n$, $R_c$, and $S_A/S_B$ are 34.5 mm, 54.5 mm, and 4. The radius of the waveguide $R$ is 14.5 mm. The STL result of numerical simulation with Eq. (2) for the given parameters agrees very well with the analytical STL. Both models indicate that the resonance frequency of HR is 870 Hz, the peak values of 53.1 dB (analytical) and 50.0 dB (numerical), respectively.

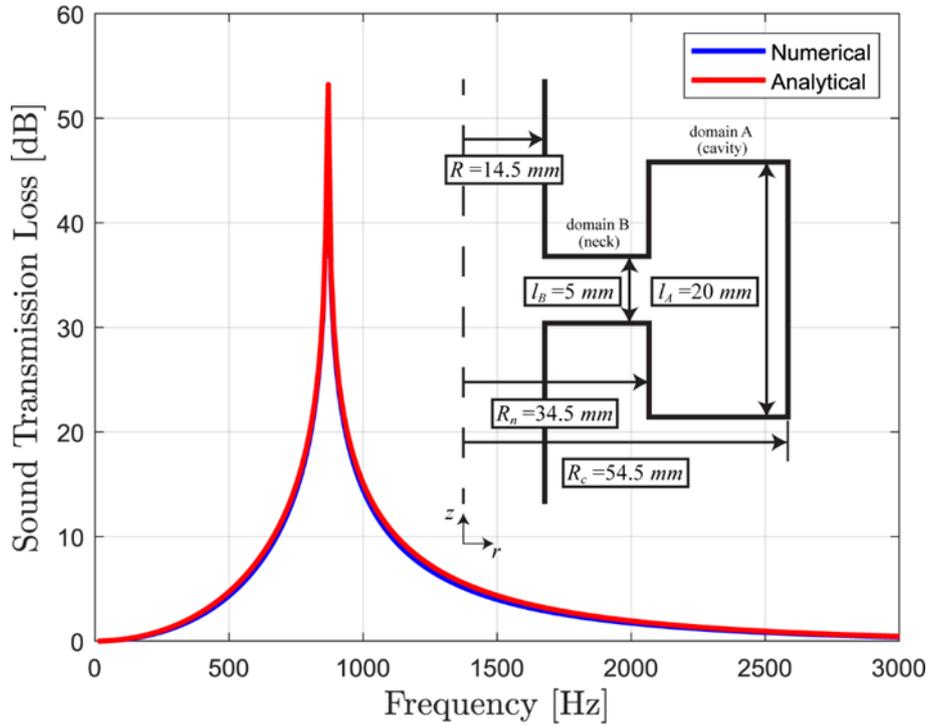

**Fig S2.** The sound transmission loss of the numerical (blue line) and analytical (red line) model

2. Detailed information about the VAE loss function

We present detailed information on the derivation of the VAE loss function. The VAE aims to generate data with a similar distribution as the input data x. To achieve this, the VAE aims at maximum likelihood estimation for parameters of marginal likelihood $p_\theta(x)$ and inference for true posterior density $p_\theta(z|x)$ that samples z generating input data x well. The Marginal likelihood $p_\theta(x)$ can be expressed as the integral of the joint distribution, considering the relationship with the latent variable z as Equation 1.

$$p_\theta(x) = \int p_\theta(x,z)dz = \int p_\theta(z)p_\theta(x|z)dz \tag{1}$$

where $p_\theta(z)$ is the prior distribution, and $p_\theta(x|z)$(decoder) is the conditional distribution of x given z. If $p_\theta(x|z)$ is a complex function like a neural network, then the true posterior density $p_\theta(z|x)$ is intractable because $p_\theta(z|x) = p_\theta(x|z)p_\theta(z)/p_\theta(x)$. Since maximum likelihood estimation like the EM algorithm is impossible, we used a variational inference by introducing a simple approximation function of the true posterior density $q_\emptyset(z|x)$ (encoder).

For the convenience of maximum likelihood estimation, the marginal likelihood is changed to the log-likelihood, and the following formula is established by Jensen's inequality.

$$\log(p_\theta(x)) = \log(\int p_\theta(x|z)p_\theta(z)dz) \geq \int \log(p_\theta(x|z))p_\theta(z)dz \tag{2}$$

For the variational inference, by introducing the approximation function $q_\emptyset(z|x)$, the equation can be rewritten as

$$\log\{p_\theta(x)\} = \log\left(\int p_\theta(x|z)\frac{p_\theta(z)}{q_\emptyset(z|x)}q_\emptyset(z|x)dz\right) \geq \int \log\left(p_\theta(x|z)\frac{p_\theta(z)}{q_\emptyset(z|x)}\right)q_\emptyset(z|x)dz$$

$$= \int \log(p_\theta(x|z))\, q_\emptyset(z|x)dz - \int \log\left(\frac{q_\emptyset(z|x)}{p_\theta(z)}\right)q_\emptyset(z|x)dz$$

$$= E_{q_\emptyset(z|x)}[\log\{p_\theta(x|z)\}] - D_{KL}(q_\emptyset(z|x)\|p_\theta(z)) \tag{3}$$

The last term in the above equation is called the evidence lower bound (ELBO). $\emptyset$ and $\theta$ are the parameters of the encoder and decoder respectively. The process of finding the parameter that maximizes the ELBO is the maximum likelihood estimation process for the marginal likelihood $p_\theta$ and also can maximize the probability of generating input x. Since gradient descent-based learning proceeds to minimize the loss function, Equation 3 can be rewritten as Equation 4.

$$L(\emptyset, \theta; x) = -E_{q_\emptyset(z|x)}[\log\{p_\theta(x|z)\}] + D_{KL}(q_\emptyset(z|x)\|p_\theta(z)) \quad (4)$$

3. Detailed network structure of the AR-VAE

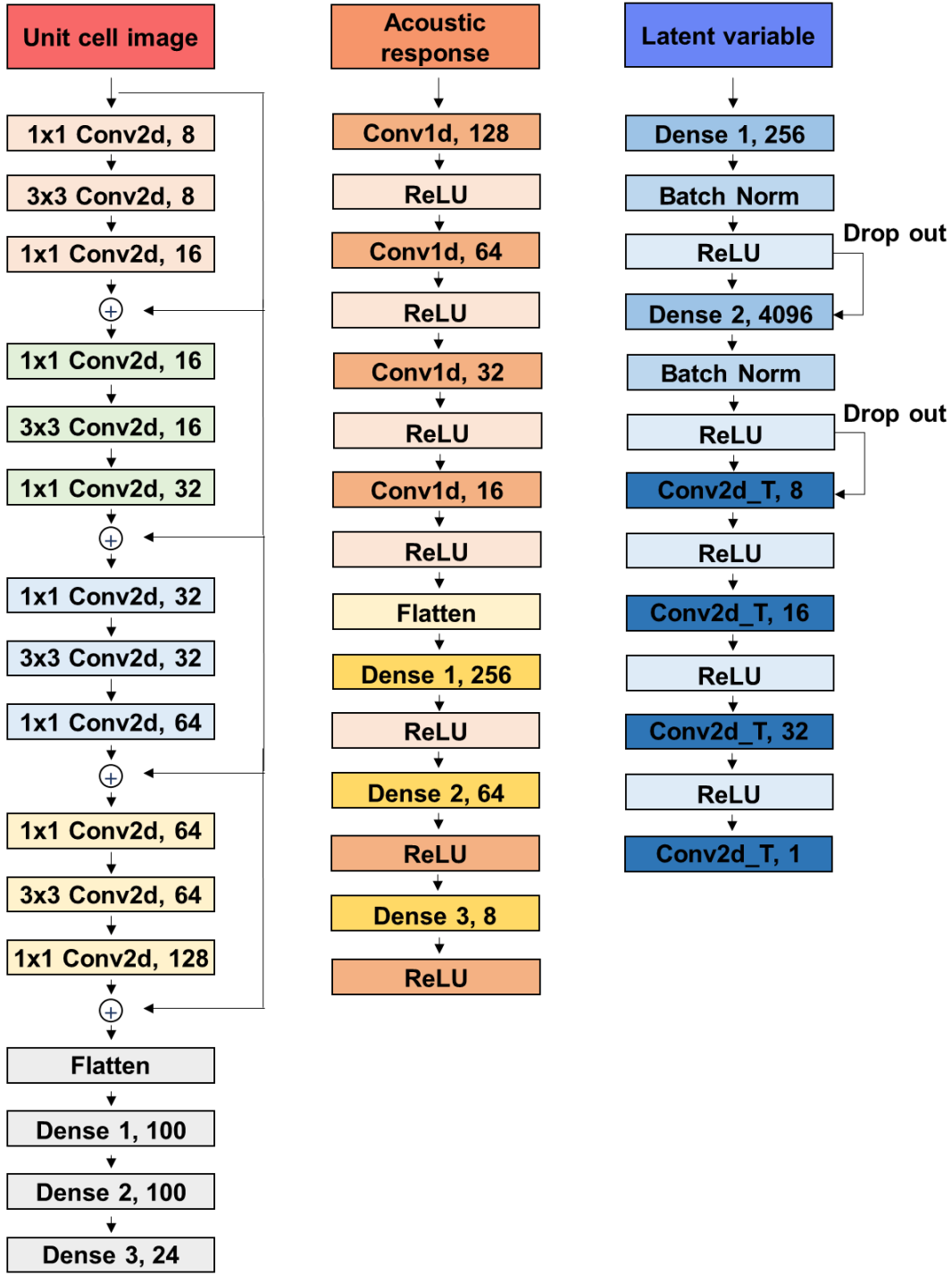

**Fig. S3** Detailed network structure of AR-VAE

4. Detailed information and results of non-parameterized VAR inverse design byAR-VAE

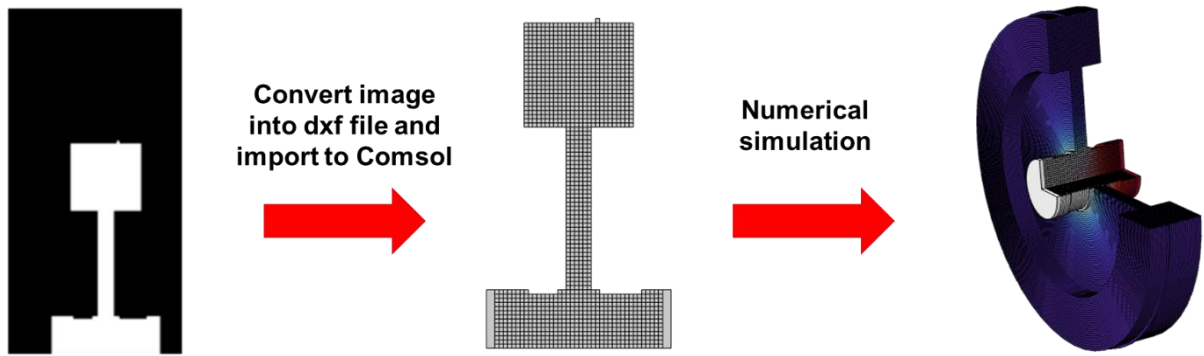

**Fig. S4** The schematic of numerical simulation for binary non-parameterized VAR cross-section images

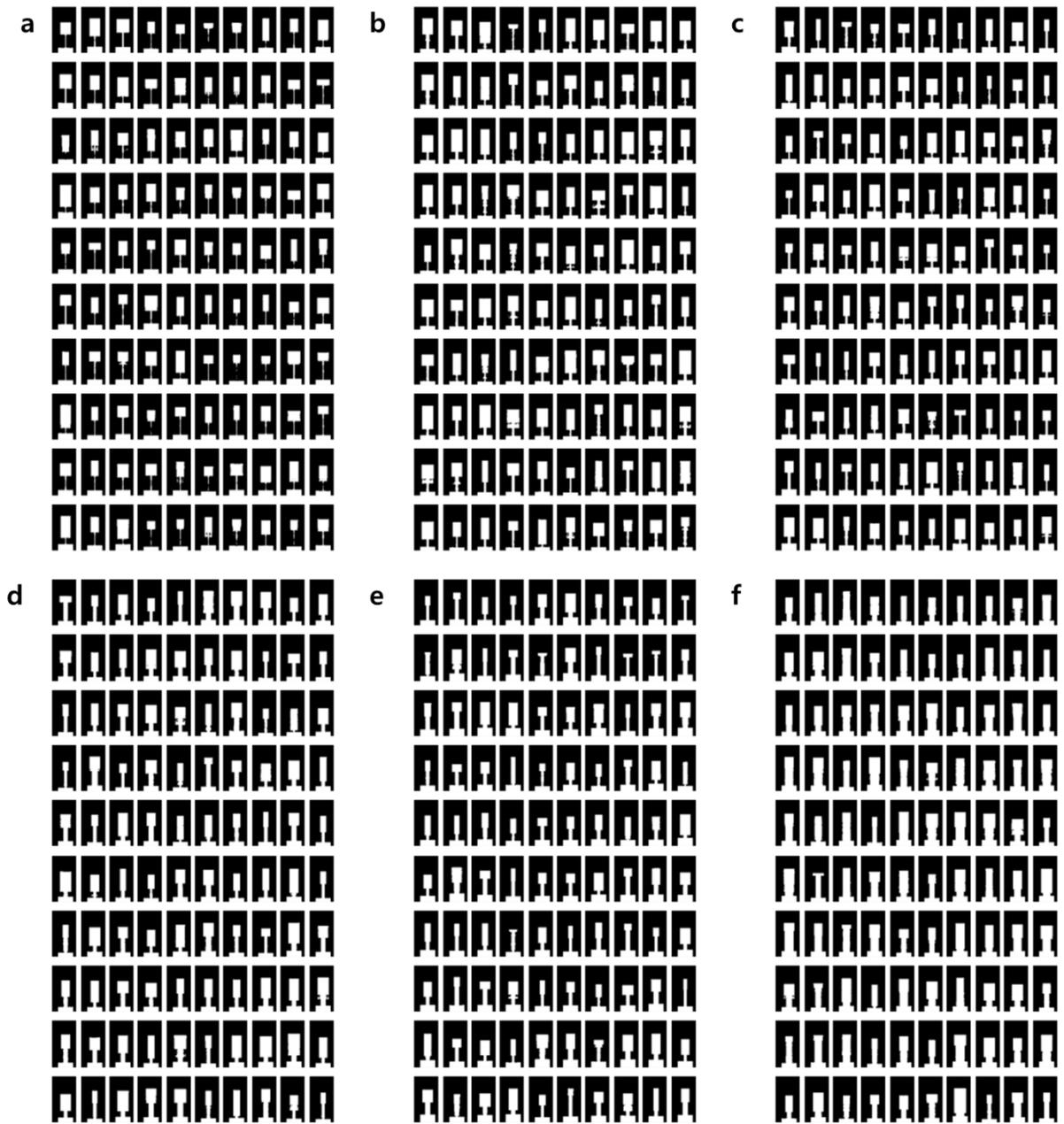

**Fig. S5** 100 sample of VAR cross-section images generated by AR-VAE for each target acoustic response at peak frequencies (a) 601 Hz, (b) 1201 Hz, (c) 801 Hz, (d) 1401 Hz, (e) 1001 Hz, and (f) 1601 Hz

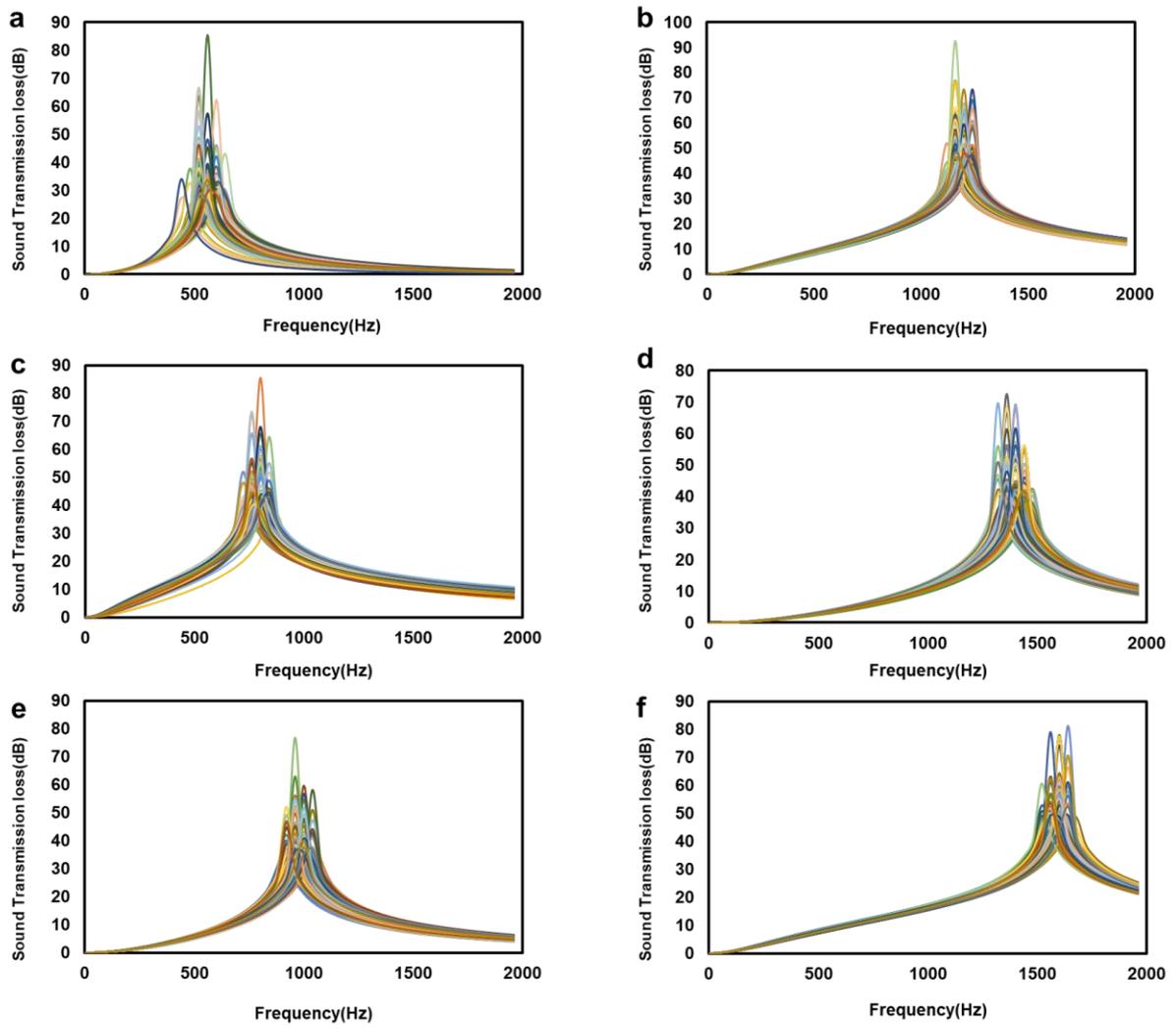

**Fig. S6** 100 sample numerical simulation results of non-parameterized VAR cross-section image for each target acoustic response at peak frequencies (a) 601 Hz, (b) 1201 Hz, (c) 801 Hz, (d) 1401 Hz, (e) 1001 Hz, and (f) 1601 Hz

5. Detailed information and results of APNN

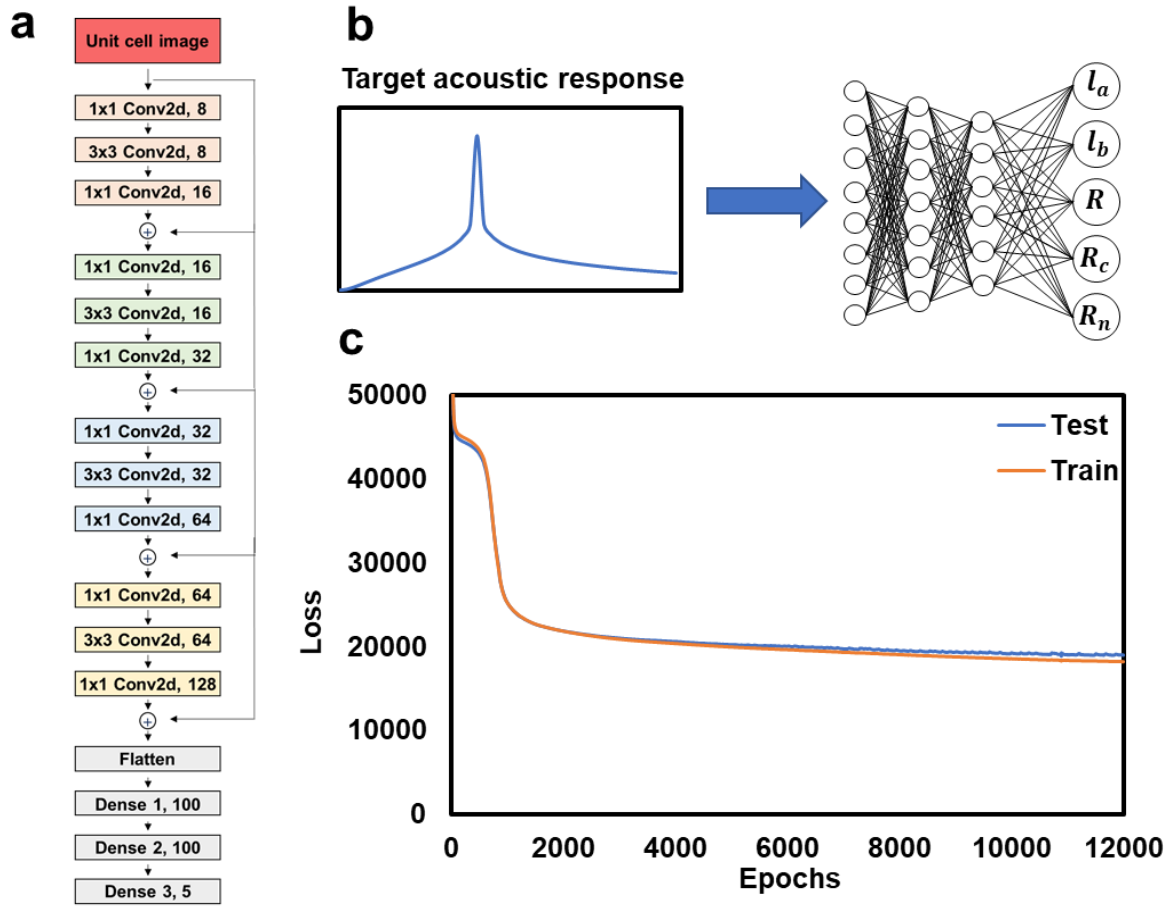

**Fig. S7** (a) The network structure of APNN, (b) The schematic of APNN, (c) The learning curve of APNN

The training process of APNN proceeded by using an Intel® Core™ i5-8500 CPU @ 3.00GHz 3.00 GHz and NVIDIA GeForce RTX3060 GPU. We set the batch size to 512, used Adam optimizer with a learning rate of 1e-5, and proceeded with training up to 12000 epochs. The training results are shown as the learning curves in Figure S7b.

6. Detailed information and results parameterized VAR cross-section image

To obtain a VAR cross-section image with a well-defined boundary, a parameter detection algorithm was developed for the inverse-designed VAR cross-section images. The horizontal length of the VAR for each row was measured by subtracting the coordinate values of starting and end pixels with a value of 1 in the horizontal axis direction. The value of $R$ was measured by counting the number of rows with the same horizontal length as $l_c$ (length of air-fluid channel). K-means clustering with two clusters was performed for the measured candidate horizontal lengths of VAR. Among the two cluster label values, the larger value was used as the $l_a$, while the smaller value was used as the $l_b$. The height of the VAR in each column was measured by subtracting the coordinate value of the starting and end pixels with a value of 1 in the vertical axis direction. Among the measured VAR heights, heights higher than the value obtained by subtracting the threshold from the largest value of the measured VAR heights were selected as candidates for the $R_c$ value. The mode of the candidates was used as the $R_c$. A VAR cross-section image without the cavity was created through $R_c$, $R$, and $l_b$. After that, a cross-section image of the cavity was acquired by subtracting the VAR cross-section image without a cavity from the inverse-designed VAR cross-section images. Through the cross-section image of the cavity, the vertical length of the cavity could be measured. $R_n$ can be calculated by subtracting the vertical length of the cavity from $R_c$. Through the above procedure, we were able to obtain the geometric parameters $R, l_b, R_n, l_a, R_c$. The parameter detection algorithm pseudo-code and results are shown in Figure s7 and the results are shown in Figure s8. We conducted a numerical simulation by constructing a 3D structure based on the geometric parameters obtained by the parameter detection algorithm.

## Parameter detection Algorithm

```
Procedure Parameter detection (image)
  for i in range(image_vertical_length):
    horizontal_length_of_ith_row= 1end_coord_of_ith_row- 1start_ coord_of_ith_row    ►measure horizontal length of each row
    if horizontal_length_of_ith_row == l_c :
      R=R+1                                                                          ►measure R
    elif horizontal_length_of_ith_row !=0 :                                           ►horizontal length candidates list
      horizontal_length_list.append(horizontal_length_of_ith_row)
  end
  kmeans_clusters=kmeans_clustering(horizontal_length_list,n_clusters=2)              ►k-means clustering to distinguish l_a, l_b
  l_a=max(kmeans_clusters)                                                            ►measure l_a
  l_b=min(kmeans_clusters)                                                            ►measure l_b
  for j in range(image_horizontal_length):
    vertical_length_of_jth_column=1end_coord_of_jth_column - 1start_ coord_of_jth_column   ►measure vertical length of each column
    vertical_length_list.append(vertical_length_of_jth_column)                        ►vertical length candidates list
  end
  for j in range(image_horizontal_length):
    if vertical_length_of_jth_column >= max(vertical_length_list)-threshold
      Rc_length_candidate_list.append(vertical_length_of_jth_column)                  ►R_c candidates list
  end
  R_c=argmax(vertical_length_candidate_list)                                          ►measure R_c
  image_without_cavity=image_generator(l_b, R, R_c)                                   ►generate image without cavity
  cavity_image=image-image_without_cavity                                             ►extract cavity image
  for j in range(cavity_image_horizontal_length):
    cavity_vertical_length_of_jth_column=1end_coord_of_jth_column - 1start_ coord_of_jth_column   ►measure vertical length of cavity
    cavity_vertical_length_list.append(cavity_vertical_length_of_jth_column)          ►Cavity vertical length candidates list
  end
  cavity_vertical_length=argmax(cavity_vertical_length_list)                          ►measure cavity vertical length
  R_n=R_c-cavity_vertical_length                                                      ►measure R_n
  return l_a, l_b, R, R_c, R_n                                                        ►return parameters
```

**Fig. S8** Pseudo-code of the parameter detection algorithm

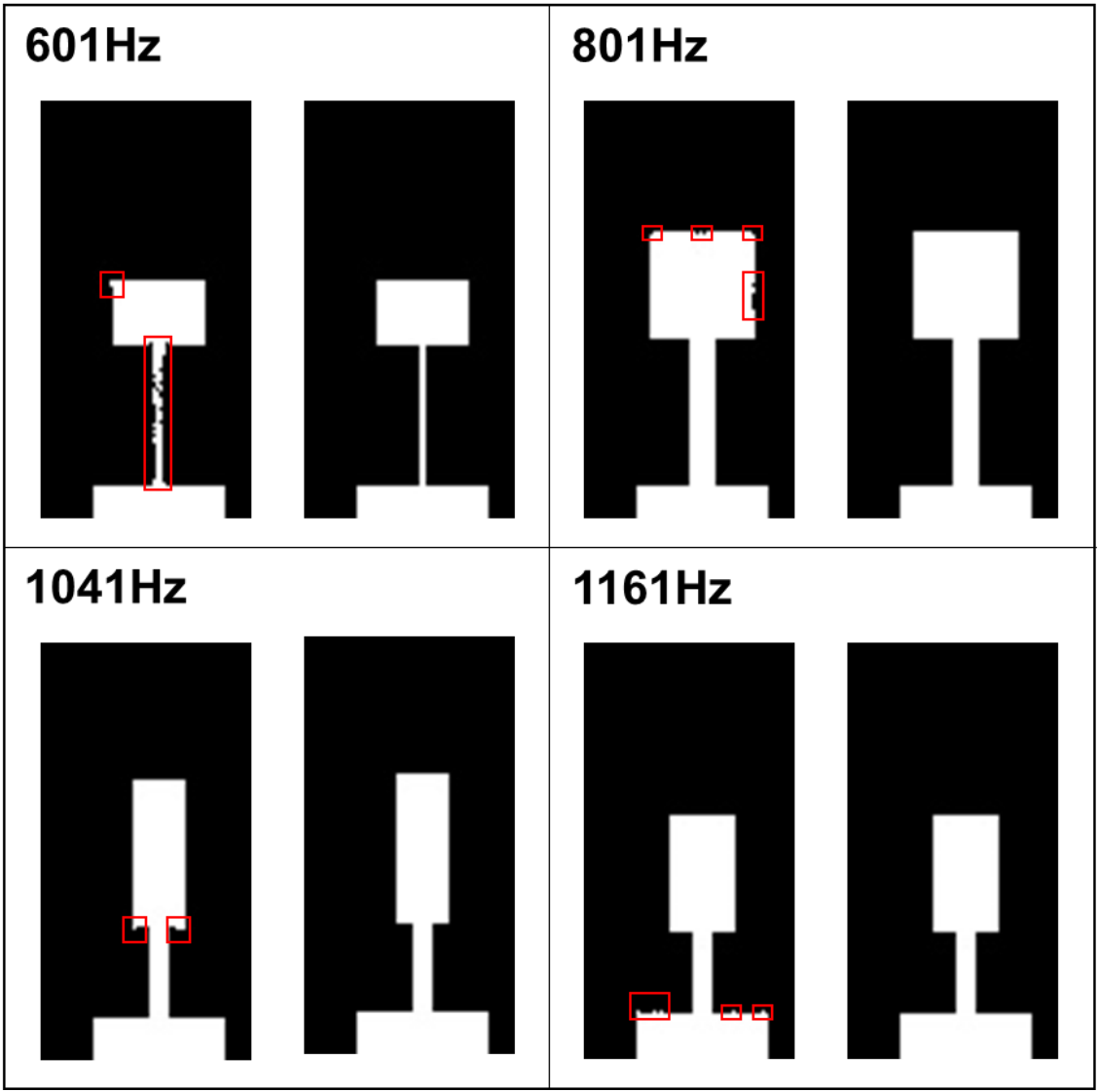

**Fig. S9** Results of parameterized VAR cross-section image

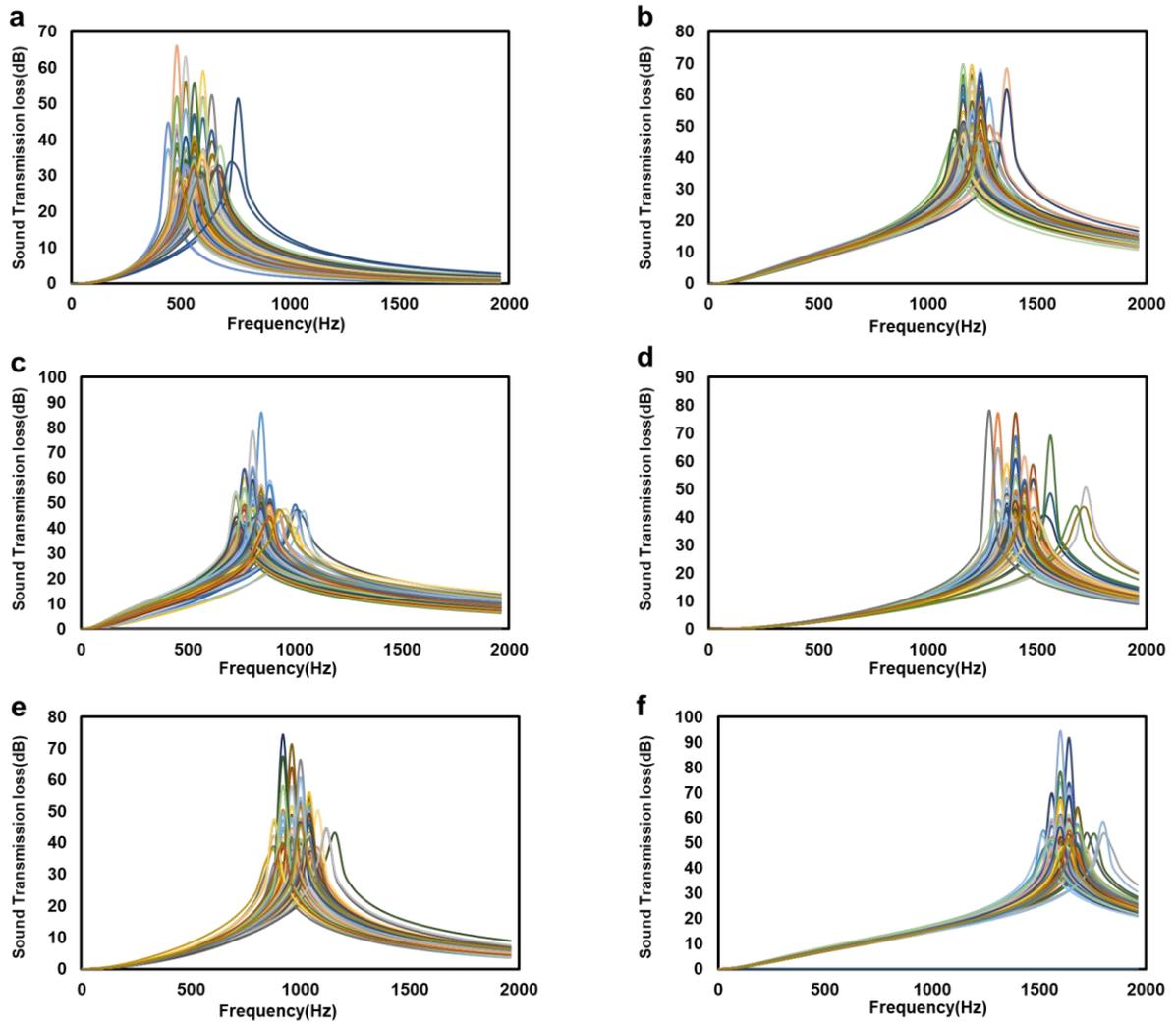

**Fig. S10** 100 sample numerical simulation results of parameterized VAR cross-section image for each target acoustic response at peak frequencies (a) 601 Hz, (b) 1201 Hz, (c) 801 Hz, (d) 1401 Hz, (e) 1001 Hz, and (f) 1601 Hz

7. Average mean squared STL error and peak frequency variance comparison results between non-parameterized inverse-designed VAR and parameterized inverse-designed VAR

**Fig. S11** 100 sample average mean squared STL error comparison for each target acoustic response

| Peak frequency | Non-parameterized inverse-designed VAR | Parameterized inverse-designed VAR |
|---|---|---|
| 601Hz | 61.785 | 62.098 |
| 801Hz | 42.676 | 69.684 |
| 1001Hz | 46.588 | 52.25 |
| 1201Hz | 43.641 | 49.608 |
| 1401Hz | 59.196 | 66.397 |
| 1601Hz | 51.514 | 80.697 |

**Fig. S12** 100 sample peak frequency variance comparison for each target acoustic response

| Peak frequency | Non-parameterized inverse-designed VAR | Parameterized inverse-designed VAR |
|---|---|---|
| 601Hz | 0.769 | 2.033 |
| 801Hz | 0.601 | 15.14 |
| 1001Hz | 0.902 | 1.85 |
| 1201Hz | 0.597 | 1.542 |
| 1401Hz | 0.96 | 3.049 |
| 1601Hz | 0.64 | 63.353 |

# Supporting Information References

[1] Nguyen H, Q Wu, X Xu, H Chen, S Tracy, and G Huang,(2020), Broadband acoustic silencer with ventilation based on slit-type Helmholtz resonators. *Applied Physics Letters,* vol. 117, no. 13, doi: 10.1063/5.0024018